\documentclass[pra,aps,superbib,citeautoscript,twocolumn,longbibliography,superscriptaddress]{revtex4-1}

\usepackage[colorlinks=true,urlcolor=blue,citecolor=blue,linkcolor=blue]{hyperref}

\usepackage{amsmath,bm}
\usepackage{amstext}
\usepackage{epsfig}
\usepackage{xcolor}
\usepackage{subfig}
\usepackage{graphicx,amsmath,amssymb,tabularx}
\usepackage{multirow}
\usepackage{array}
\usepackage{dsfont}
\usepackage{caption}
\usepackage{cleveref}
\usepackage{ulem}
\usepackage[toc]{appendix}
\usepackage{physics}
\usepackage{mathrsfs}  
\usepackage{amsbsy}
\usepackage{float}
\definecolor{dgreen}{rgb}{0,.5,0}
\definecolor{dred}{rgb}{.7,.0,.0}
\definecolor{dblue}{RGB}{11, 139, 230}

\DeclareMathAlphabet\mathbfcal{OMS}{cmsy}{b}{n}

\newcommand{\bxi}{\bm{\xi}}

\newcommand{\br}{\mathbf{r}}

\newcommand{\ie}{{\it i.e.}}

\newcommand{\be}{\begin{eqnarray}}
\newcommand{\ee}{\end{eqnarray}}
\newcommand{\bse}{\begin{subequations}}
\newcommand{\ese}{\end{subequations}}
\DeclareMathAlphabet\mathbfcal{OMS}{cmsy}{b}{n}

\newcommand{\rev}[1]{{{#1}}}

\begin{document}

\title
  {Charged excitations made neutral: $N$-centered ensemble density functional theory of Fukui functions
   }
\author{Lucien Dupuy}
\affiliation{Laboratoire de Chimie Quantique,
Institut de Chimie, CNRS / Universit\'{e} de Strasbourg,
4 rue Blaise Pascal, 67000 Strasbourg, France}
\affiliation{University of Strasbourg Institute for Advanced Study,
5, all\'{e}e du G\'{e}n\'{e}ral Rouvillois, F-67083 Strasbourg, France}
\email{lucien.dupuy@unistra.fr}
\author{Emmanuel Fromager}
\affiliation{Laboratoire de Chimie Quantique,
Institut de Chimie, CNRS / Universit\'{e} de Strasbourg,
4 rue Blaise Pascal, 67000 Strasbourg, France}
\affiliation{University of Strasbourg Institute for Advanced Study,
5, all\'{e}e du G\'{e}n\'{e}ral Rouvillois, F-67083 Strasbourg, France}

\begin{abstract}
An in-principle exact working equation to compute electronic affinity and ionization Fukui functions is derived within the $N$-centered (Nc) ensemble extension of density functional theory (DFT). It circumvents the kernel derivative discontinuity problem of DFT for fractional electron numbers, whose contribution is recovered through weight derivatives of the ensemble density functional potential. Thus, it allows for the design of alternative and effective approximations, such as the weight-dependent scaling of regular functionals or the interpolation between known limits of Nc ensembles.  

\end{abstract}

\maketitle

{\it Introduction} --
No matter how much we succeed in making accurate computer simulations of chemical reactions routine, there will always be a need for chemical descriptors of reactivity. While they cannot systematically replace simulation nor experiment, their appeal is to bring "primitive patterns of understanding"\footnote{Attributed to C.A. Coulson by R. McWeeny in the preface of McWeeny R (1979) Coulson's valence. Oxford University Press, Oxford} reducing the over-abundant information generated in Ab Initio resolution of so many processes to key principles we can grasp. In a time where accurate molecular simulations are not so routinely feasible, their guidance is as valuable at the prediction stage than for ensuing rationalization.

As chemistry and condensed-matter physics is first and foremost dictated by the response of electrons to perturbation (from other electrons and nuclei, light, applied electric field and many more), they are the focus of properties descriptors. The advent of Density Functional Theory (DFT), an in-principle exact framework in which every property of the system follows from the electronic density, brought a solid foundation from which sharp definitions of insightful concepts -- such as electronegativity~\cite{parr1980density} and hardness~\cite{parr1983absolute} -- emerged. Further analysis of increasingly local responses of the electronic energy to changes in the external potential and number of electrons bloomed into the field of Conceptual Density Functional Theory~\cite{CDFTrev2003,CDFTrev2008} (CDFT). In turn, their computation from corresponding functional derivatives of the (to-be-approximated) universal functional proved instrumental in understanding what information practical Density Functional Approximations (DFA) should incorporate: By focusing efforts on reproducing more relevant properties, DFA design can (hopefully) find shortcuts on the meandering path towards the exact functional~\cite{gouldfukui2017}. This is a contribution of the present work.

We will focus more specifically on the Fukui function, introduced by Parr and Yang~\cite{parrFukui84} as a tool to understand the relative reactivity of different sites in a molecule~\cite{ayers2000perspective}. It is defined as the derivative of the electronic density with respect to the electron number, a somewhat puzzling concept for isolated molecules~\cite{Bader_chempot82} which must have an integer number of electrons, but very natural for atoms or fragments within a bigger molecular arrangement resulting in partial charges on each subunit~\cite{AiM_Bader90}. Based on the reasoning that an approaching reagent will prefer the site most favorable to addition/removal of an electron for a nucleophilic/electrophilic attack, the corresponding Fukui function will reach a higher value the more reactive a site is~\cite{ayers2000perspective}. It offers a (sometimes critical~\cite{ayers_fukui_05}) improvement over frontier molecular orbital theory~\cite{fukui1952molecular,fmo1954theory,fmo82} by inclusion of orbital relaxation effects. 

Interpretation and computation of derivatives with respect to the number of electrons analytically, {\it i.e.} connecting them to the $N$-electron Kohn Sham (KS) system of DFT and the Hartree-exchange-correlation (Hxc) density functional, is much more insightful (and more accurate) than their finite-differences evaluation from separate calculations on the $N$ and $N-1$ electron systems~\cite{Weitao_fukui_2012}. To that end, extension of DFT to fractional electron numbers by Perdew, Parr, Levy and Balduz (PPLB) showed the generalized energy functional possesses derivative discontinuities when crossing an integer number of electrons~\cite{perdew1982density}, a behavior which in turn affects properties obtained through differentiating it~\cite{perdew1982density,mori2008localization,cohen2008fractional,cohen2008insights,stein2010fundamental,zheng2011improving,cohen2011challenges,perdew2017understanding}. Accurate predictions depend on inclusion of discontinuous shifts of the Hxc fields at integer electron number thresholds, which practical (local and semi-local) DFAs lack. The most well-known case is the missed contribution of Hxc potential derivative discontinuities resulting in systematic underestimation of fundamental gaps in solids~\cite{perdew2017understanding}, also relevant for molecular dissociation and charge transfer~\cite{hodgson2017interatomic,Kraisler21_From}. It is less widely appreciated that similar discontinuities also appear in the Hxc kernel, responsible for aforementioned orbital relaxation effects, making computation of the Fukui function within PPLB challenging~\cite{Hellgren12_Effect}. Constraints on KS orbital densities arising from piecewise-linearity requirement of the physical density were recently investigated~\cite{Kraisler_piecewiselin_dens_25}, aiming to identify and correct density-driven errors in DFT calculations. The highest occupied molecular orbital (HOMO) dependence on its fractional occupation proved to be intricate, invalidating the frozen approximation and even linear behavior in simple atomic systems. In some cases, it was found to not even be two-point-Taylor-expandable, with preliminary analysis suggesting the divergence of Taylor coefficients stems from approximate functionals being used.

The recently introduced $N$-centered extension of ensemble density functional theory~\cite{senjean2018unified,PRA21_Hodgson_exact_Nc-eDFT_1D,Cernatic2022} (Nc EDFT) is an alternative exact approach to charged electronic excitations that shed a new light on the discontinuity problem. While PPLB considers an ensemble of $N$ and $N-1$ electrons pure states with relative weights constraining the obtained ensemble density, the Nc ensemble density always integrates to the central number of electrons $N$ by construction, no matter the weights ({\it vide infra}). The ensemble density and weights become independent variables of the theory, which transform the elusive energy derivative discontinuity contributions into weight-derivatives of the Hxc ensemble functional. This offers a promising way to "recycle" (semi-) local DFAs lacking derivative discontinuities~\cite{loos2020weightdependent}. It also provides a considerable gain in scope as it can describe any charged or neutral excitation within a unified picture~\cite{Cernatic2024_Neutral}, and can be adapted to treat open systems~\cite{Senjean_2020}. The theory can also benefit from progress made in recent years in ensemble DFA development~\cite{PRL19_Gould_DD_correlation,Fromager_2020,PRL20_Gould_Hartree_def_from_ACDF_th,Gould2023_Electronic,gould2024local,gould2026ensemblization,Gould2025_PRL_tate-Specific}. Moreover, the very recent derivation by the authors of exact expressions for neutrally-excited state densities~\cite{Fromager2025indvElevel} and their responses~\cite{dupuy2025_exact_static} within an ensemble makes computation of properties beyond excitation energies a timely pursuit.

In this Letter, we derive an exact $N$-centered EDFT of Fukui functions and discuss the fundamental and practical implications of our findings in the description of second (and possibly higher)-order density-functional Hxc derivatives. We then investigate two strategies to build a practical EDFA and study their performance on the Hubbard dimer model system~\cite{carrascal2015hubbard} as proof of principle. First, we will show how dressing the regular ground-state Hxc functional with a weight-dependent scaling function allows to capture the weight-derivative term. Then, we will demonstrate the performance of an EDFA built by interpolating between known low and high correlation limits of the Hxc ensemble density functional.

\bigskip
{\it Fukui functions from Nc EDFT} - Nc ensembles consist of a reference $N$-electron ground ($\nu=0$) and $N_\nu$-electron excited ($\nu>0$) states $\{\ket{\Psi_\nu}\}$ which are all eigenstates of a given electronic Hamiltonian $\hat{H}=\hat{T}+\hat{W}_{\rm ee}+\int d\br\, v(\br)\hat{n}(\br)$, where $\hat{T}$ and $\hat{W}_{\rm ee}$ are the electronic kinetic and repulsion energy operators, respectively, while $v(\br)$ is the external local potential at position $\br$ and $\hat{n}(\br)$ is the one-electron density operator at that same position. Note that, in this context, the ground state of the $N_\nu$-electron system in which $N_\nu\neq N_0=N$ is described as a (charged) excited state with respect to $\Psi_0$. Mathematically, an ensemble is described by a density matrix operator $\hat{\Gamma}^{\bxi}=\sum_{\nu\geq 0}\xi_\nu \ket{\Psi_\nu} \bra{\Psi_\nu}$, $\xi_\nu\geq 0$ being the (arbitrarily chosen for now) weights that characterize the ensemble. In the particular case of Nc ensembles, $\hat{\Gamma}^{\bxi}$ 
reads more explicitly as follows,
\begin{equation}
    \hat{\Gamma}^{\bxi} = \left(1-\sum_{\lambda>0} \frac{N_{\lambda}}{N} \xi_\lambda\right) \ket{\Psi_0} \bra{\Psi_0} + \sum_{\lambda>0} \xi_\lambda \ket{\Psi_\lambda} \bra{\Psi_\lambda}
    .
\end{equation}
It is {\it not} normalized (\ie, $\sum_{\nu\geq 0}\xi_\nu\neq 1$, in general) on purpose~\cite{Cernatic2024_Neutral}, for reasons that will become clearer in the following. In fact, the expression of $\xi_0$ in terms of the excited-state weights $\bxi = \{\xi_{\lambda}\}_{\lambda>0}$ (monotonically decreasing with the energy for all states describing the same number of electrons~\cite{gross1988rayleigh}) ensures that, by construction, the Nc ensemble electronic density $n^{\bxi}({\bf r}):=\Tr\left[\hat{\Gamma}^{\bxi}\hat{n}(\br)\right]$, where $\Tr$ denotes the trace, or, more explicitly,  
\begin{equation}\label{eq:true_int_Nc_ens_dens}
    n^{\bxi}({\bf r}) = \left(1-\sum_{\lambda>0} \frac{N_{\lambda}}{N} \xi_\lambda\right) n_{\Psi_{0}}({\bf r}) + \sum_{\lambda>0} \xi_\lambda n_{\Psi_{\lambda}} ({\bf r}),
\end{equation}
where $n_{\Psi}(\br):=\langle \Psi\vert \hat{n}(\br)\vert \Psi\rangle$, integrates to the reference (so-called central) electron number $N$ of the reference $\nu=0$ ground state. It thus generalizes the Theophilou-Gross-Oliveira-Kohn (TGOK) ensemble framework for neutral excitations~\cite{JPC79_Theophilou_equi-ensembles,gross1988rayleigh, gross1988density,oliveira1988density} to also describe charged excitations, and has already been explored in some detail~\cite{senjean2018unified,deur2019ground,PRA21_Hodgson_exact_Nc-eDFT_1D,Cernatic2022,cernatic2024extended_doubles,Cernatic2024_Neutral}, in particular considering consequences of an ensemble density $n$ and $\bxi$ being independent variables of the theory, unlike in regular (PPLB) DFT~\cite{Cernatic2022}. In the conventional KS formulation of Nc EDFT, the ensemble density is mapped onto a non-interacting density matrix operator $\hat{\Gamma}_{s}^{\bxi}=\left(1-\sum_{\lambda>0} \frac{N_{\lambda}}{N} \xi_\lambda\right) \vert{\Phi^{\bxi}_{0}}\rangle \langle{\Phi^{\bxi}_{0}}\vert + \sum_{\lambda>0} \xi_\lambda \vert{\Phi^{\bxi}_{\lambda}}\rangle \langle{\Phi^{\bxi}_{\lambda}}\vert$, \ie,
\be\label{eq:KS_Nc_ens_dens_mapping}
\Tr\left[\hat{\Gamma}_{s}^{\bxi}\hat{n}(\br)\right]=\Tr\left[\hat{\Gamma}^{\bxi}\hat{n}(\br)\right]
=n^{\bxi}({\bf r}),
\ee
where the (weight-dependent) non-interacting KS states $\left\{\ket{\Phi^{\bxi}_{\nu}}\right\}_{\nu\geq 0}$ are eigenstates of the ensemble KS Hamiltonian $\hat{H}_{s}^{\bxi}=\hat{T}+\int d\br (v(\br)+v^{\bxi}_{\rm Hxc}[n^{\bxi}](\br))\hat{n}(\br)$, $v^{\bxi}_{\rm Hxc}[n](\br)\equiv \delta E^{\bxi}_{\rm Hxc}[n]/\delta n(\br)$ being the analog for Nc ensembles of the Hxc density-functional potential. By exploiting the linear variation in Eq.~(\ref{eq:true_int_Nc_ens_dens}) of the ensemble density with respect to the weights we can trivially and exactly extract, by analogy with TGOK-DFT~\cite{Fromager2025indvElevel,dupuy2025_exact_static},  any individual-state density from the Nc ensemble one as follows,
$
n_{\Psi_\nu}=\frac{N_\nu}{N}n^{\bxi}+\sum_{\lambda>0}( \delta_{\lambda\nu}-\frac{N_{\nu}}{N} \xi_\lambda)\frac{\partial n^{\bxi}}{\partial \xi_\lambda}$. According to the KS mapping of Eq.~(\ref{eq:KS_Nc_ens_dens_mapping}), it leads to the more explicit expression (where we dropped the $\br$ dependence, for the sake of compactness),
\be\label{eq:ind_dens_from_KS_rsp_func}
n_{\Psi_{\nu}}=n_{\Phi^{\bxi}_{\nu}}+\sum_{\lambda>0} \Big( \delta_{\lambda\nu}-\frac{N_{\nu}}{N} \xi_\lambda\Big)\chi^{\bxi}_s \star\dfrac{\partial}{\partial \xi_\lambda}\left(v^{\bxi}_{\rm Hxc}[n^{\bxi}]\right),
\ee
where we use the shorthand notation $\chi\star f: \br \mapsto \int d\br' \chi(\br,\br')f(\br')$, and $\chi^{\bxi}_s$ is the non-interacting KS analog of the Nc ensemble density-density linear response function $\chi^{\bxi}:\,(\br,\br')\mapsto \chi^{\bxi}(\br,\br')=\delta n^{\bxi}(\br)/\delta v(\br')$. Note that the two response functions are related through the following Dyson equation,    
\begin{equation} \label{eq:ens_dyson}
    \big(\chi^{\bxi}\big)^{-1} = \big(\chi^{\bxi}_s\big)^{-1} - f^{\bxi}_{\rm Hxc},
\end{equation}    
where $f^{\bxi}_{\rm Hxc}=f^{\bxi}_{\rm Hxc}[n^{\bxi}]$ is the Nc ensemble density-functional Hxc kernel $f^{\bxi}_{\rm Hxc}[n](\br,\br')\equiv \delta v^{\bxi}_{\rm Hxc}[n](\br)/\delta n(\br')$ evaluated at the exact ensemble density. If we further expand and rewrite Eq.~(\ref{eq:ind_dens_from_KS_rsp_func}) as follows, 
\be
\begin{split}
&\big(\chi^{\bxi}_s\big)^{-1}\star \left( n_{\Psi_{\nu}}-n_{\Phi^{\bxi}_{\nu}}\right)
= f^{\bxi}_{\rm Hxc}\star \left(n_{\Psi_{\nu}}-\frac{N_\nu}{N}n^{\bxi}\right)
\\
&+\sum_{\lambda>0} \Big( \delta_{\lambda\nu}-\frac{N_{\nu}}{N} \xi_\lambda\Big) \frac{\partial v_{\rm Hxc}^{\bxi}[n]}{\partial \xi_{\lambda}}\Big|_{n=n^{\bxi}}
,  
\end{split}
\ee

or, equivalently,
\be
\begin{split}
\big(\chi^{\bxi}\big)^{-1}\star n_{\Psi_{\nu}}
&=\big(\chi^{\bxi}_s\big)^{-1}n_{\Phi^{\bxi}_{\nu}}- \frac{N_\nu}{N}f^{\bxi}_{\rm Hxc}\star n^{\bxi}
\\
&\quad+\sum_{\lambda>0} \Big( \delta_{\lambda\nu}-\frac{N_{\nu}}{N} \xi_\lambda\Big) \frac{\partial v_{\rm Hxc}^{\bxi}[n]}{\partial \xi_{\lambda}}\Big|_{n=n^{\bxi}},
\end{split}
\ee
we finally obtain, by noticing that $\chi^{\bxi}\big(\chi^{\bxi}_s\big)^{-1}=1+\chi^{\bxi}\star f^{\bxi}_{\rm Hxc}$, the following exact and compact relation between the true and KS individual-state densities, 
\begin{equation} \label{eq:ind_dens}
    \begin{split}
        n_{\Psi_{\nu}}  \underset{\nu \geq 0}{=} &\big(1+\chi^{\bxi} \star f_{\rm Hxc}^{\bxi}\big) \star n_{\Phi^{\bxi}_{\nu}} - \frac{N_{\nu}}{N}\chi^{\bxi} \star f_{\rm Hxc}^{\bxi} \star n^{\bxi}
        \\
        & + \chi^{\bxi} \star \sum_{\lambda>0} \Big( \delta_{\lambda\nu}-\frac{N_{\nu}}{N} \xi_\lambda\Big) \frac{\partial v_{\rm Hxc}^{\bxi}[n]}{\partial \xi_{\lambda}}\Big|_{n=n^{\bxi}}.
    \end{split}
\end{equation}
Note that Eq.~\eqref{eq:ind_dens} generalizes Eq.~(104) of Ref.~\citenum{dupuy2025_exact_static} to charged excited states. As it applies universally to any $N_\nu$-electron ground or excited state ($N_\nu=N,N\pm 1,N\pm 2,\ldots$), it can be (straightforwardly) applied to the calculation of physical (interacting) Fukui functions $f_{\kappa}={\theta}_{\kappa}( n_{\Psi_\kappa} - n_{\Psi_0})$, for which $N_\kappa\neq N$ and ${\theta}_{\kappa}={\rm sgn}(N_\kappa-N)$, thus leading to the following exact relation between $f_{\kappa}$ and its non-interacting KS analog $f^{\bxi}_{s,\kappa}={\theta}_{\kappa}( n_{\Phi^{\bxi}_\kappa} - n_{\Phi^{\bxi}_0})$,     
\begin{equation} \label{eq:fukui}
    \begin{split}
        f_{\kappa} 
        &=  \big(1+\chi^{\bxi} \star f_{\rm Hxc}^{\bxi}\big) \star f^{\bxi}_{s,\kappa} - \abs{N_{\kappa}-N} \; \chi^{\bxi} \star f_{\rm Hxc}^{\bxi} \star \frac{n^{\bxi}}{N} 
        \\
        &\quad  + {\theta}_{\kappa} \; \chi^{\bxi} \star  \sum_{\lambda> 0} \Big( \delta_{\lambda\kappa}-\frac{N_{\kappa}-N}{N} \xi_{\lambda}  \Big) \frac{\partial v_{\rm Hxc}^{\bxi}[n]}{\partial \xi_{\lambda}}\Big|_{n=n^{\bxi}}. 
    \end{split}
\end{equation} 
Eq.~\eqref{eq:fukui} is the first key result of this Letter.

Interestingly, in regular (PPLB) DFT, this relation reduces to the first term on the right-hand side (the second and third terms simply do not exist in PPLB), which is then dependent on the aforementioned discontinuous Hxc kernel for accurate predictions~\cite{Hellgren12_Effect}. It appears that, within the Nc EDFT formalism, the impact of discontinuities on the calculation of Fukui functions is described explicitly through ensemble weight derivatives, as readily seen from the last term on the right-hand side of Eq.~\eqref{eq:fukui}. Thus, we generalize a feature that was already known for the evaluation of ionization potentials and electron affinities~\cite{Cernatic2024_Neutral}. This equivalence can be made even more transparent by using the invariance of Eq.~\eqref{eq:fukui} under symmetric shifts in the ensemble Hxc kernel,
\begin{equation}
    f_{\rm Hxc}^{\bxi} ({\bf r},{\bf r}') \rightarrow f_{\rm Hxc}^{\bxi} ({\bf r},{\bf r}') + g({\bf r}) + g({\bf r}'),
\end{equation}
since $\int d\br\, f^{\bxi}_{s,\kappa}(\br)=\abs{N_{\kappa}-N}$ and $\int d\br\, \chi^{\bxi}(\br,\br')=\delta \left(\int d\br\, n^{\bxi}(\br)\right)/\delta v(\br')=\delta N/\delta v(\br')$=0. This fundamental property is specific to Nc EDFT. Indeed, in PPLB, physically-relevant variations of the density lead to a change of particle number, thus breaking the gauge-invariance~\cite{Hellgren12_Effect}. Note that, within Nc EDFT, we can always make a gauge choice for a specific charged excitation $[0\rightarrow \kappa]$ such that the last two terms on the right-hand side of Eq.~(\ref{eq:fukui}) cancel out, \ie, 
\begin{equation} \label{eq:kernel_shift}
    \begin{split}
            & \chi^{\bxi} \star \bigg[ \frac{|N_{\kappa}-N|}{N} f_{\rm Hxc}^{\bxi [0\rightarrow \kappa]} \star n^{\bxi} 
            \\& - {\theta}_{\kappa} \sum_{\lambda> 0} \Big( \delta_{\lambda\kappa}-\frac{N_{\kappa}-N}{N} \xi_{\lambda}  \Big) \frac{\partial v_{\rm Hxc}^{\bxi}[n]}{\partial \xi_{\lambda}}\Big|_{n=n^{\bxi}} \bigg] = 0,        
    \end{split}
\end{equation}
so that Eq.~(\ref{eq:fukui}) becomes formally identical to that of PPLB, with {\it no} approximations made. Therefore, each excitation process requires applying a specific ($\kappa$-dependent) weight-dependent symmetric shift $g^{\bxi [0\rightarrow \kappa]}({\bf r})+g^{\bxi [0\rightarrow \kappa]}({\bf r}')=f_{\rm Hxc}^{\bxi [0\rightarrow \kappa]}(\br,\br')-f_{\rm Hxc}^{\bxi}(\br,\br')$ to the ensemble Hxc kernel to ensure a systematic exactification of the PPLB formula, which then holds not only for ground but also for {\it excited} $N_\kappa$-electron states. According to Eq.~(\ref{eq:kernel_shift}),  
\begin{equation}\label{eq:shift_general_expression}
\begin{split}
    g^{\bxi [0\rightarrow \kappa]}
    &=\dfrac{1}{N_{\kappa}-N}\frac{\partial v_{\rm Hxc}^{\bxi}[n]}{\partial \xi_{\kappa}}\Big|_{n=n^{\bxi}}
\\
&
\quad -\dfrac{1}{N}\left(\sum_{\lambda> 0}\xi_{\lambda}\frac{\partial v_{\rm Hxc}^{\bxi}[n]}{\partial \xi_{\lambda}}\Big|_{n=n^{\bxi}}\right) - f_{\rm Hxc}^{\bxi} \star \dfrac{n^{\bxi}}{N}.    
\end{split}
\end{equation}
Eq.~(\ref{eq:shift_general_expression}), which is our second and central formal result, establishes the equivalence between the discontinuities in the Hxc kernel and the weight derivatives of the ensemble Hxc density-functional potential (see the first term on the right-hand side). The clarity of the Nc EDFT picture is made manifest in the fact that no assumption about the asymptotic behavior of the density needs to be invoked~\cite{Hellgren12_Effect} to rationalize and evaluate the shift. This is relevant for analysis of periodic systems, where Fukui functions can rationalize surface reactivity~\cite{fukui_func_solid25}, for example.

\medskip
{\it Practical strategies for Nc EDFA design} -- The opportunity to employ regular (ground-state) DFAs in this context arises from noticing that Eq.~\eqref{eq:fukui} holds even in the zero-weight limit of the theory, where the ensemble Hxc kernel, response function and density become those of regular $N$-electron ground-state DFT. If we focus more specifically on the ground-state theory of a single electron addition ($\kappa\equiv +$) or removal ($\kappa\equiv -$), it reads 
\begin{equation} \label{eq:fukui0}
    \begin{split}
        f_{\pm} &=   \big(1+\chi \star f_{\rm Hxc}\big) \star f_{s,\pm} -\; \chi \star f_{\rm Hxc} \star \frac{n_0}{N}
        \\
        &\quad  \pm \; \chi \star \frac{\partial v_{\rm Hxc}^{\bxi}[n]}{\partial \xi_{\pm}}\Big|_{n=n_0}^{\bxi=0}, 
    \end{split}
\end{equation}
$f_{s,+}$ being the lowest unoccupied KS molecular orbital (LUMO) in the $N$-electron ground state, $f_{s,-}$ the HOMO, and $n_0:=n^{\bxi=0}=n_{\Psi_0}$. This is an astounding simplification where the only piece needed (aside from the regular $N$-electron Hxc kernel) to compute Fukui functions, in principle exactly, is the weight-derivatives of the Nc ensemble density-functional Hxc potential evaluated at $\bxi\equiv \{\xi_+,\xi_-\}=0$. 
\rev{To that aim, an appealing strategy is to ``dress'' Hx and correlation parts of a regular (weight-independent) DFA with weight-dependent scaling functionals, \ie,
\begin{equation} \label{eq:pot_scaling}
     E^{\bxi}_{\rm Hxc}[n] \approx s_{\rm Hx}(\bxi) \; E_{\rm Hx}[n] + s_{\rm c}[\bxi,n] \; E_{\rm c}[n],
\end{equation}
where $s_{\rm Hx}(\bxi=0)=s_{\rm c}[\bxi=0,n]=1$, by construction. 
\rev{The above ansatz, which can be rationalized from exact uniform coordinate scaling relations~\cite{Toulouse2023,Nagy_ensAC,Scott2024_Exact}, becomes more transparent in both high- and low-density limits, as shown in the supplemental material~\cite{suppmat}. In particular, the irreducible, non-trivial density dependence of the correlation scaling becomes clear in the corresponding weakly and strongly correlated regimes, at the Nc-EDFT level of generality\cite{mirtschink2013derivative}.} 
\rev{According to Eq.~(\ref{eq:pot_scaling}), the weight-derivative of the Hxc density-functional potential reads
\be\label{eq:most_general_form_scaled_EDFA_weight_deriv_pot}
\begin{split}
    \dfrac{\partial v^{\bxi}_{\rm Hxc}[n](\br)}{\partial \xi_\pm}\approx &  
\frac{\partial s_{\rm Hx}(\bxi)}{\partial \xi_{\pm}}v_{\rm Hx}[n](\br)
+\frac{\partial s_{\rm c}[\bxi,n]}{\partial \xi_{\pm}}v_{\rm c}[n](\br)
\\
& +E_{\rm c}[n]\dfrac{\delta}{\delta n(\br)} \left[\frac{\partial s_{\rm c}[\bxi,n]}{\partial \xi_{\pm}}\right],
\end{split}
\ee
which offers great flexibility to reproduce system's properties accurately~\cite{suppmat}.}
Indeed, in the spirit of hybrid range-separated (RSH) DFT~\cite{Morgante2023_Strategies}, one can optimally tune (OT) the Hx scaling function to
satisfy the ionization potential theorem of Nc EDFT~\cite{senjean2018unified,Cernatic2024_Neutral,suppmat}, 
\begin{equation} \label{eq:IP_general}
    \begin{split}
        I^{N+\frac{1}{2}\pm\frac{1}{2}} = & - \varepsilon^{\bxi}_{N+\frac{1}{2}\pm\frac{1}{2}} - \frac{E^{\bxi}_{\rm Hxc}[n^{\bxi}] - \left( v^{\bxi} \middle| n^{\bxi} \right)}{N} 
        \\
        & +\left(\frac{\xi_{\pm}}{N}\mp 1\right) \frac{\partial E^{\bxi}_{\rm Hxc}[n]}{\partial \xi_{\pm}}\Big|_{n=n^{\bxi}},
    \end{split}
\end{equation}
along the lines of Ref.~\onlinecite{dupuy2026ensembledensityfunctionaltheory}. \rev{The weight-derivative of the correlation scaling functional can then be adjusted so that the resulting ensemble scaled DFA reproduces any desired (possibly highly accurate) Fukui function~\cite{suppmat}, with the second line of Eq.~\eqref{eq:most_general_form_scaled_EDFA_weight_deriv_pot} bringing the necessary {$\br$}-dependence (see Eq.~(\ref{eq:fukui0})). Crucially, the ansatz of Eq.~(\ref{eq:pot_scaling}) is in principle flexible enough to overcome shortcomings of OT-RSH strategies~\cite{Morgante2023_Strategies} using only a (comparatively simpler) semi-local DFA as basic ingredient, making it appealing as a framework for efficient construction of ensemble Hxc functionals by deep learning~\cite{skala1.1}, for example}. As detailed in the supplemental material~\cite{suppmat}, the procedure can also be used in a fully iterative and self-contained way (\ie, without any training data),  provided the full range of weight values is explored at each iteration. While this strategy lays the foundations of an alternative approach to EDFA developments, its implementation would require further simplifications, like numerical interpolations~\cite{rath2025interpolating} between weights, for example, so that it can be turned, ultimately, into a practical computational tool.}

\medskip
{\it \rev{Exact solution and approximations for a model system}
} -- As a proof of principle, we apply the present Nc EDFT formalism to the simple but nontrivial asymmetric Hubbard dimer~\cite{carrascal2015hubbard,carrascal2016corrigendum,Fromager_2020,deur2017exact}. The model allows to study all regimes of electronic correlation, by tuning the on-site (atomic) electronic repulsion to the electronic hopping (which is analogous to the kinetic energy strength) ratio $U/t$, for a given difference $\Delta v$ in external potential between the two sites, while the ensemble Hxc functional and its derivatives can be computed exactly~\cite{suppmat}. We focus on charged excitations between different electronic ground states, as described in Eq.~(\ref{eq:fukui0}), with the central number of electrons set to $N=2$. Exact ensemble densities are systematically used so that we can focus on functional-driven (and, in particular, weight-driven) errors. 
\rev{We define the density as the occupation of site 0~\cite{suppmat}, and we compute Fukui functions corresponding to that convention.} 
\rev{For our purpose, we apply (equivalently) the scaling strategy of Eq.~\eqref{eq:pot_scaling} directly to the Hxc potential, \ie, we now define $s_{\rm Hx/c}[\bxi,n]:={v^{\bxi}_{\rm Hx/c}[n]}/{v_{\rm Hx/c}[n]}$~\cite{suppmat}.}
\rev{We start with the (analytically solvable in the model~\cite{suppmat}) ensemble exact exchange-only (EEXX) approximation, which has been used extensively to compute excitation energies~\cite{yang2014exact,pribramjones2014excitations,yang2017direct,gould2018charge,sagredo2018can,deur2018exploring,senjean2018unified,deur2019ground,Senjean_2020}. 
}

\begin{figure}[h!]
    \centering
    \includegraphics[width=\linewidth]{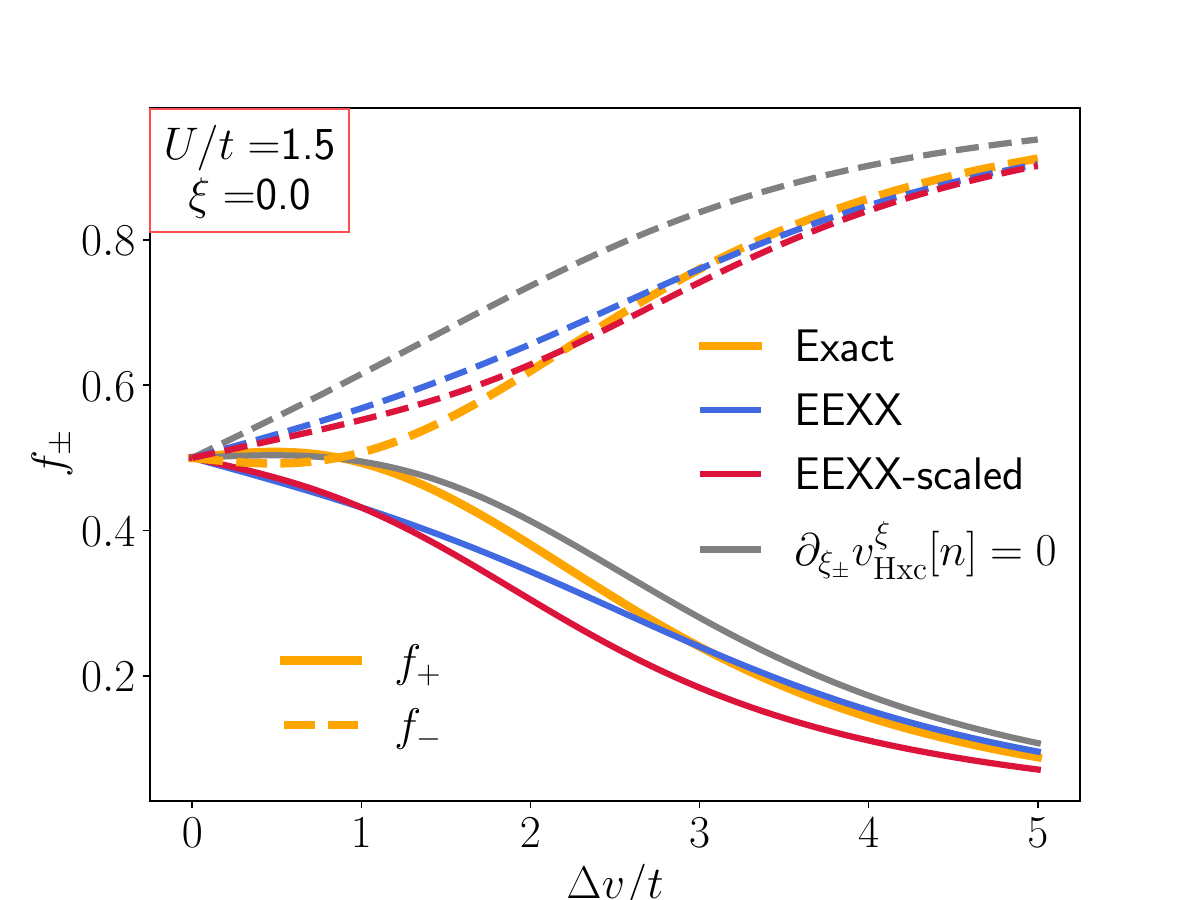}
    \caption{
    Performance of the EEXX-scaled EDFA in evaluating Fukui functions (as functions of the external asymmetry potential $\Delta v$) from the zero-weight limit of Nc EDFT in the moderately correlated $U/t=1.5$ regime  of the Hubbard dimer.
    }
    \label{fig:onescalevsnwU1.5x0}
    \end{figure}

\begin{figure}[h!]
    \centering
    \includegraphics[width=\linewidth]{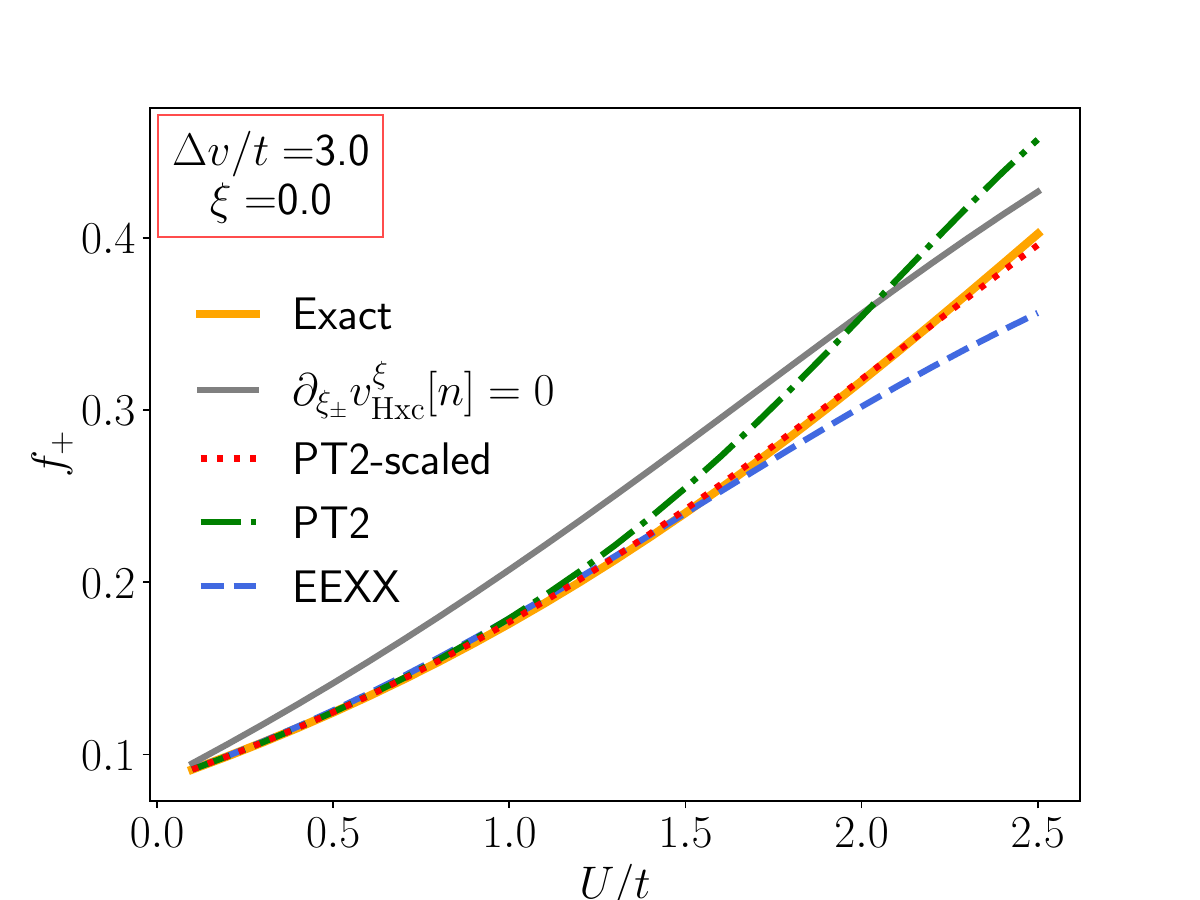}
    \caption{Performance of the PT2-scaled EDFA in evaluating the affinity Fukui function (as a function of the interaction strength $U$) up to the stronger correlation $U/t=2.5$ regime of the asymmetric ($\Delta v/t=3.0$) Hubbard dimer.
    }
    \label{fig:fpPT2scalingDv3}
\end{figure}

\rev{As shown in Fig.~\ref{fig:onescalevsnwU1.5x0} in the moderately correlated $U/t=1.5$ regime, it improves substantially over the complete neglect of weight-derivatives in the evaluation of the ionization Fukui function $f_-$, but not for the affinity one ($f_+$). In the latter case, ensemble correlation effects mostly compensate those of EEXX. The importance of using a specific, weight- and density-dependent scaling for the correlation~\cite{suppmat}, is also illustrated by results using the EEXX-scaled full ground-state Hxc density-functional potential $v^{\bxi}_{\rm Hxc}[n] \approx s_{\rm Hx}(\bxi)v_{\rm Hxc}[n]$, which worsens significantly the prediction of $f_+$ while bringing only a modest improvement for $f_-$.}
\rev{To obtain a proper correlation scaling, the extension to ensembles~\cite{Yang2021_Second} of G\"{o}rling--Levy perturbation theory~\cite{gorling1993correlation, gorling1994exact} through second order (PT2) can be used, its derivation being presented in the supplementary material~\cite{suppmat}. In itself, PT2 (applied on top of EEXX) fixes the flaws of EEXX when $U/t=1.5$~\cite{suppmat}. But more importantly, using the PT2-derived correlation scaling functional in our ansatz extends the applicability of PT2 (through the scaling functionals) to stronger correlation regimes, with a remarkably high accuracy, as illustrated in Fig.~\ref{fig:fpPT2scalingDv3} for the affinity Fukui function. A deeper analysis of the correlation potential, which illustrates how the scaling approach corrects erroneous behaviors of PT2 for strongly asymmetric dimers, is provided in the supplemental material~\cite{suppmat}.}


For completeness, we also extended to charged excitations the \rev{strategy of interpolating between known exact limits of weakly-correlated and strictly-correlated electrons from} Ref.~\citenum{Gould2023_Electronic}. In the Hubbard dimer, it translates into the following Pad\'e approximant (see the supplementary material for the complete expression~\cite{suppmat}) to the ensemble density-functional Hxc energy,
\begin{equation} \label{eq:pade1_main_text}
    E_{\rm Hxc}^{\bxi}[n] \approx a^{\bxi}[n] \, U +  \frac{b^{\bxi}[n] \, U^2}{1+\frac{b^{\bxi}[n]}{\gamma^{\bxi}[n]-a^{\bxi}[n]}U},
\end{equation}
which is exact in both weakly ($U/t\rightarrow 0$) and strictly correlated ($U/t\rightarrow +\infty$) limits, by construction, and from which the Fukui functions can be evaluated through density and weight derivatives. To be applicable (and numerically stable) to intermediate correlation regimes, a smoothed version of the coefficient $\gamma^{\bxi}[n]$ in Eq.~(\ref{eq:pade1_main_text}) has been designed such that the Hxc kernel is reproduced at all interaction strengths in the pure ground-state ($\bxi=0$) and symmetric ($\Delta v =0$) case~\cite{suppmat}. The latter is the analog of the uniform electron gas in this context, echoing recent work to develop (semi-)local density functionals for excited states~\cite{Loos2025excstUEG}.    

\begin{figure}[h!]
    \centering
    \includegraphics[width=\linewidth]{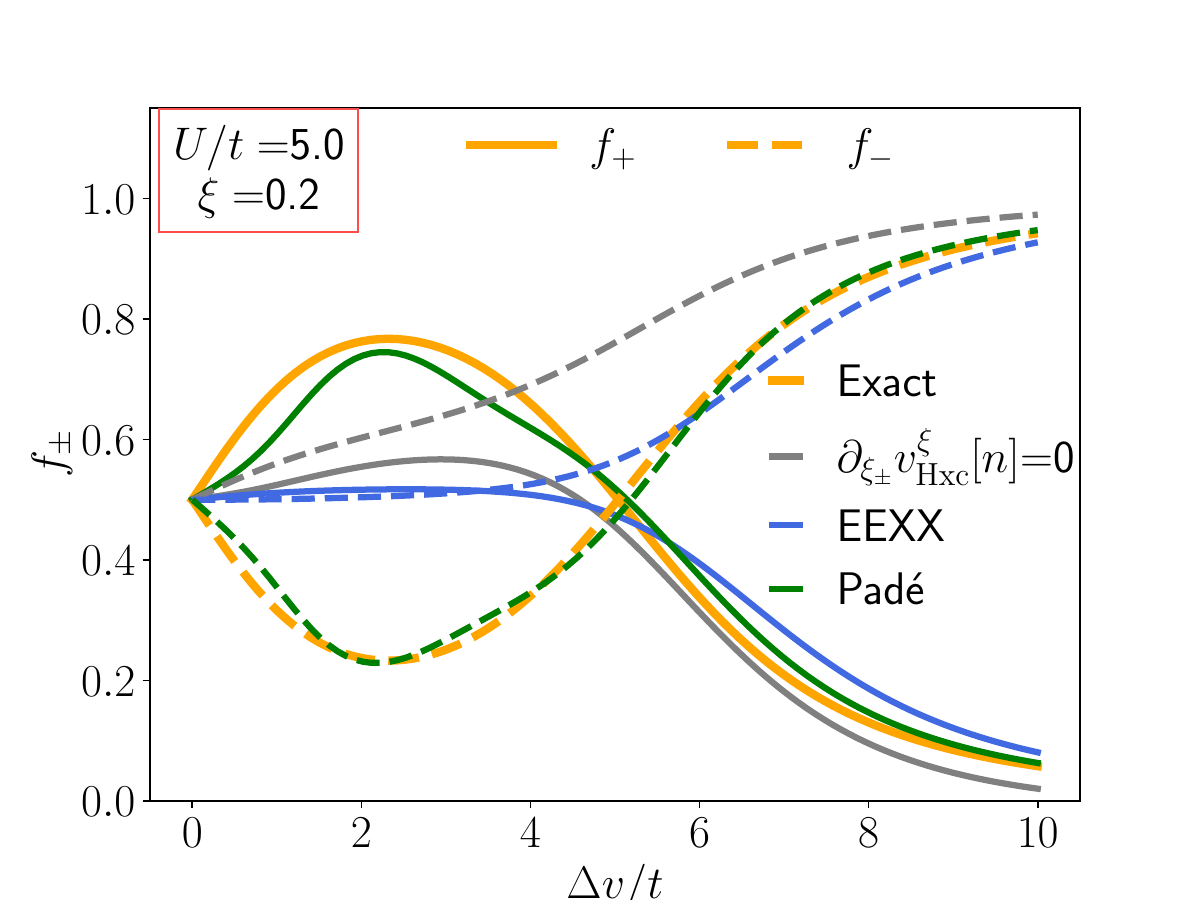}
    \caption{Same as Fig.~\ref{fig:onescalevsnwU1.5x0} for the Pad\'{e} approximant-based EDFA in the strongly correlated $U/t=5$ regime and away from the zero-weight limit ($\xi_+=\xi_-=0.2$). 
    }
    \label{fig:padesvsnwU5x0.2}
\end{figure}
The robustness of the smoothed Pad\'{e} approximant away from the zero-weight limit it learned from (we take $\xi_+=\xi_-=\xi = 0.2$) is illustrated in Fig.~\ref{fig:padesvsnwU5x0.2}, where it successfully applies to the strongly correlated $U/t=5$ dimer in {\it all} asymmetry regimes, unlike EEXX, as expected.

In summary, an alternative ($N$-centered) EDFT of Fukui functions has been derived, thus revealing that important but challenging higher-order Hxc derivative discontinuities can be tackled, in principle exactly, through ensemble weight derivative corrections to regular DFT. As a result, significant progress made recently in the understanding and application of EDFT~\cite{gould2026ensemblization,Gould2025_PRL_tate-Specific} can now be exploited for computing accurately charged excited-state properties beyond the energy, as illustrated in this Letter.

This work has benefited from support provided by the University of Strasbourg Institute for Advanced Study (USIAS) for a Fellowship, within the French national programme “Investment for the future” (IdEx-Unistra).

\bibliography{biblio}

\onecolumngrid

\appendix

\section*{Appendix}

\tableofcontents

\section{Construction of {\it ab initio} scaled EDFAs}

\subsection{Context}

We want to develop the following (so-called scaled) $N$-centered ensemble density-functional ansatz,
\be\label{eqapp:scaled_EDFA_ansatz}
E^{\bxi}_{\rm Hxc}[n]\equiv s_{\rm Hx}[\bxi,n]E_{\rm Hx}[n]+s_{\rm c}[\bxi,n]E_{\rm c}[n], 
\ee
where $E_{\rm Hx}[n]=E_{\rm Hx}^{\bxi=0}[n]$ and $E_{\rm c}[n]=E_{\rm c}^{\bxi=0}[n]$ are the regular Hx and correlation functionals, respectively. In practice, the latter can be described at the (semi)-local level of density-functional approximation (DFA), thus allowing for the description of density-functional derivative discontinuity contributions to energy gaps and Fukui functions at a relatively low computational cost, through a proper incorporation of ensemble weight dependencies into the Hx and correlation scaling functionals ($s_{\rm Hx}[\bxi,n]$ and $s_{\rm c}[\bxi,n]$, respectively), as discussed in further detail in Sec.~\ref{sec:learning_in_zero_weight_limit}. But first we want to identify properties of the latter scaling density functionals from the exact uniform coordinate scaling relations (see Ref.~\citenum{Toulouse2023} for a comprehensive review on the topic) that are fulfilled not only by pure ground states but also by $N$-centered density-functional ensembles~\cite{dupuy2026ensembledensityfunctionaltheory}, as discussed further in the following.\\

Note that, depending on the context, {\it scaling} can refer either to the uniform coordinate scaling or to the scaling of regular Hx and correlation DFAs, as implemented in Eq.~(\ref{eqapp:scaled_EDFA_ansatz}). These are of course two (completely) different types of scaling.  

\subsection{Exact uniform coordinate scaling relations}

We denote $\left\{\Phi_\nu^{\bxi}[n]\right\}_{\nu\geq 0}$ the normalized ($N$-centered) ensemble density-functional KS wavefunctions. As readily seen from the non-interacting limit of Eqs.~(A4) and (A7) in Ref.~\citenum{dupuy2026ensembledensityfunctionaltheory}, and the fact that
\be
\int d\br_1\int d\br_2\ldots \int d\br_{N_\nu}\left\vert\Phi_\nu^{\bxi}[n](\gamma\br_1,\gamma\br_2,\ldots,\gamma \br_{N_\nu})\right\vert^2=\dfrac{1}{\gamma^{3N_\nu}}
\ee
and
\be
N_\nu\gamma^{3N_\nu}\int d\br_2\ldots \int d\br_{N_\nu}\left\vert\Phi_\nu^{\bxi}[n](\gamma\br,\gamma\br_2,\ldots,\gamma \br_{N_\nu})\right\vert^2=\gamma^3 n_{\Phi_\nu^{\bxi}[n]}(\gamma\br),
\ee
we have
\be
\gamma^{\frac{3N_\nu}{2}}\Phi_\nu^{\bxi}[n](\gamma\br_1,\gamma\br_2,\ldots,\gamma \br_{N_\nu})=\Phi_\nu^{\bxi}[n_\gamma](\br_1,\br_2,\ldots,\br_{N_\nu}),
\ee
where $n_\gamma(\br)=\gamma^3 n(\gamma\br)$ is the ensemble density obtained from $n$ through the uniform coordinate scaling $\br \rightarrow \gamma \br$. As a result, the ensemble Hx energy evaluated for the scaled density reads~\cite{dupuy2026ensembledensityfunctionaltheory} 
\be\label{eqapp:simp_ens_Hx_func_scaled_dens}
\begin{split}
E^{\bxi}_{\rm Hx}[n_\gamma]
&=\sum_{\nu\geq 0}\xi_\nu \langle \Phi_\nu^{\bxi}[n_\gamma]\vert\hat{W}_{\rm ee}\vert \Phi_\nu^{\bxi}[n_\gamma]\rangle
\\
&=
\sum_{\nu\geq 0}\xi_\nu \gamma^{3N_\nu}\sum^{N_\nu}_{1\leq i<j}\int d\br_1\int d\br_2\ldots \int d\br_{N_\nu}\dfrac{\left\vert\Phi_\nu^{\bxi}[n](\gamma\br_1,\gamma\br_2,\ldots,\gamma \br_{N_\nu})\right\vert^2}{\vert\br_i-\br_j \vert}
\\
&=\gamma\sum_{\nu\geq 0}\xi_\nu\langle \Phi_\nu^{\bxi}[n]\vert\hat{W}_{\rm ee}\vert \Phi_\nu^{\bxi}[n]\rangle
\\
&=\gamma E_{\rm Hx}^{\bxi}[n],
\end{split}
\ee
or, equivalently, from the substitution $\gamma\rightarrow 1/\gamma$,
\be\label{eqapp:exact_scaling_constraint_Nc_ens_Hx}
E_{\rm Hx}^{\bxi}[n]=\gamma E^{\bxi}_{\rm Hx}[n_{1/\gamma}].
\ee
Therefore, if we consider the ensemble Levy--Lieb density functional with the scaled interaction $\gamma \hat{W}_{\rm ee}$, \ie,
\be
F^{{\bxi},\gamma} [n]= \underset{\hat{\gamma}^{\bxi}\rightarrow n}{\rm min}  {\rm Tr} \left[ \hat{\gamma}^{\bxi} (\hat{T}+\gamma\hat{W}_{\rm ee}) \right],
\ee
where $\hat{\gamma}^{\bxi}$ is a trial $N$-centered ensemble density matrix operator and the density constraint reads $\Tr\left[\hat{\gamma}^{\bxi}\hat{n}(\br)\right]=n(\br)$,
and then apply the exact scaling relations~\cite{dupuy2026ensembledensityfunctionaltheory}
\be\label{eqapp:exact_scaling_relation_Fxi}
F^{{\bxi},\gamma} [n]=\gamma^2 F^{{\bxi},\gamma=1} [n_{1/\gamma}]=\gamma^2 F^{{\bxi}} [n_{1/\gamma}], 
\ee
\be\label{eqapp:exact_scaling_relation_Tsxi}
F^{{\bxi},\gamma=0} [n]=T_{\rm s}^{{\bxi}} [n]=\gamma^2 T_{\rm s}^{{\bxi}} [n_{1/\gamma}],
\ee
and
\be
E^{\bxi,\gamma}_{\rm Hx}[n]=\sum_{\nu\geq 0}\xi_\nu\langle \Phi_\nu^{\bxi}[n]\vert\gamma\hat{W}_{\rm ee}\vert \Phi_\nu^{\bxi}[n]\rangle=\gamma E_{\rm Hx}^{\bxi}[n]=\gamma^2 E^{\bxi}_{\rm Hx}[n_{1/\gamma}],
\ee
it comes 
\be
\begin{split}
E^{\bxi,\gamma}_{\rm c}[n]&=F^{{\bxi},\gamma} [n] - T_{\rm s}^{{\bxi}} [n] - E^{\bxi,\gamma}_{\rm Hx}[n] 
\\
&=\gamma^2\left( F^{{\bxi}} [n_{1/\gamma}]-T_{\rm s}^{{\bxi}} [n_{1/\gamma}]-E^{\bxi}_{\rm Hx}[n_{1/\gamma}]\right),
\end{split}
\ee
thus leading to the final relation
\be\label{eqapp:exact_constraint_Nc_ens_corr_func}
E^{\bxi,\gamma}_{\rm c}[n]=\gamma^2 E^{\bxi}_{\rm c}[n_{1/\gamma}].
\ee
In the following we will exploit the exact constraints of Eqs.~(\ref{eqapp:exact_scaling_constraint_Nc_ens_Hx}) and (\ref{eqapp:exact_constraint_Nc_ens_corr_func}) to learn more about the Hx and correlation scaling density functionals (that have been introduced in Eq.~(\ref{eqapp:scaled_EDFA_ansatz})), respectively.  

\subsection{Exact properties of the Hx scaling functional}

Let us consider the following general form of the scaled (in the sense of Eq.~(\ref{eqapp:scaled_EDFA_ansatz})) ensemble ansatz for the exact $N$-centered ensemble Hx functional $E^{\bxi,\gamma}_{\rm Hx}[n]$, where the two-electron repulsion is scaled by $\gamma$, \ie,
\be
E^{\bxi,\gamma}_{\rm Hx}[n]\equiv s^{\gamma}_{\rm Hx}[\bxi,n]E^{\gamma}_{\rm Hx}[n],  
\ee
where $E^{\gamma}_{\rm Hx}[n]=\gamma E_{\rm Hx}[n]$. First of all, from the exact relation $E^{\bxi,\gamma}_{\rm Hx}[n]=\gamma E_{\rm Hx}^{\bxi}[n]$ it immediately follows that
\be
s^{\gamma}_{\rm Hx}[\bxi,n]E^{\gamma}_{\rm Hx}[n]=\gamma s^{\gamma}_{\rm Hx}[\bxi,n]E_{\rm Hx}[n]=\gamma s_{\rm Hx}[\bxi,n]E_{\rm Hx}[n],
\ee
where $s_{\rm Hx}[\bxi,n]=s^{\gamma=1}_{\rm Hx}[\bxi,n]$, which implies the $\gamma${\it -independence} of the Hx scaling functional:
\be
s^{\gamma}_{\rm Hx}[\bxi,n]=s_{\rm Hx}[\bxi,n],\;\forall \gamma>0.
\ee
Moreover, the second exact constraint of Eq.~(\ref{eqapp:exact_scaling_constraint_Nc_ens_Hx}) [which holds for any $\bxi$ and, therefore, in the regular ground-state $\bxi=0$ limit too] implies  
\be
s_{\rm Hx}[\bxi,n]E_{\rm Hx}[n]=\gamma s_{\rm Hx}[\bxi,n]E_{\rm Hx}[n_{1/\gamma}]=\gamma s_{\rm Hx}[\bxi,n_{1/\gamma}]E_{\rm Hx}[n_{1/\gamma}], 
\ee
thus leading to the exact scaling relation,
\be\label{eqapp:exact_constraint_scalingHx_func}
s_{\rm Hx}[\bxi,n]=s_{\rm Hx}[\bxi,n_{1/\gamma}],\;\forall \gamma>0,
\ee
which will have important implications. Indeed, if we search for a simple ensemble local density approximation (ELDA) such as
\be
s_{\rm Hx}[\bxi,n]\overset{\rm ELDA}{\equiv} \dfrac{s_{\rm Hx}(\bxi)}{\int d\br\;n(\br)} \int d\br\;\left(n(\br)\right)^\beta,
\ee
by analogy with the exchange functional of the uniform electron gas, for example, the constraint of Eq.~(\ref{eqapp:exact_constraint_scalingHx_func}) implies
\be
\int d\br\;\left(n(\br)\right)^\beta=\int d\br\;\left(n_{1/\gamma}(\br)\right)^\beta=\gamma^{-3(\beta-1)}\int \dfrac{d\br}{\gamma^3}\;\left(n(\br/\gamma)\right)^\beta,\;\forall \gamma>0,
\ee
thus leading to $\beta=1$ and the density-{\it independence} of the Hx scaling functional, which becomes a simple {\it function} of the ensemble weights:
\be\label{eqapp:simplification_sx_no_dens_dep}
s_{\rm Hx}[\bxi,n]\overset{\rm ELDA}{\equiv} s_{\rm Hx}(\bxi).
\ee
Regarding the calculation of Fukui functions, the key quantity to model is the weight-dependent ensemble density-functional Hx {\it potential}, which is therefore obtained at this level of approximation by a simple weight-dependent scaling of the regular (ground-state) Hx density-functional potential, \ie,
\be
v^{\bxi}_{\rm Hx}[n]\approx s_{\rm Hx}(\bxi) v_{\rm Hx}[n]. 
\ee
Interestingly, the exact ensemble Hx potential of the Hubbard dimer model that we study in detail in the Letter exhibits the exact same property (see Eq.~(\ref{eqapp:ens_ehx})).

\subsection{Exact properties of the correlation scaling  functional}

We now consider the following general form of the scaled (in the sense of Eq.~(\ref{eqapp:scaled_EDFA_ansatz})) ensemble ansatz for the exact $N$-centered ensemble correlation functional $E^{\bxi,\gamma}_{\rm c}[n]$, where the two-electron repulsion is scaled by $\gamma$, \ie,
\be
E^{\bxi,\gamma}_{\rm c}[n]\equiv s^{\gamma}_{\rm c}[\bxi,n]E^{\gamma}_{\rm c}[n].
\ee
The exact constraint of Eq.~(\ref{eqapp:exact_constraint_Nc_ens_corr_func}) [which holds for any $\bxi$ and, therefore, in the regular ground-state $\bxi=0$ limit too] leads to the following relation that the correlation scaling functional should fulfill:
\be
s^{\gamma}_{\rm c}[\bxi,n]E^{\gamma}_{\rm c}[n]=\gamma^2 s^{\gamma}_{\rm c}[\bxi,n]E_{\rm c}[n_{1/\gamma}]=\gamma^2 s_{\rm c}[\bxi,n_{1/\gamma}]E_{\rm c}[n_{1/\gamma}],
\ee
where $s_{\rm c}[\bxi,n]=s^{\gamma=1}_{\rm c}[\bxi,n]$,
or, equivalently,
\be
s^{\gamma}_{\rm c}[\bxi,n]=s_{\rm c}[\bxi,n_{1/\gamma}],\;\forall \gamma>0.
\ee

The above equation allows us to get further insight into the correlation scaling functional, by considering the high- and low-density limiting cases~\cite{Toulouse2023}. In the high-density case ($\gamma\rightarrow 0$ in $n_{1/\gamma}$), we have, according to Eq.~(\ref{eqapp:exact_constraint_Nc_ens_corr_func}),   
\be
E^{\bxi}_{\rm c}[n_{1/\gamma}]=\dfrac{1}{\gamma^2}E^{\bxi,\gamma}_{\rm c}[n]
\underset{\gamma\rightarrow 0}{\rightarrow} E^{\bxi,{\rm GL2}}_{\rm c}[n], 
\ee
where $E^{\bxi,{\rm GL2}}_{\rm c}[n]$ is the extension to $N$-centered ensembles of the density-functional correlation energy through second order in G\"{o}rling--Levy perturbation theory (GL2) for Theophilou--Gross-Oliveira--Kohn (TGOK) ensembles of ground and neutrally excited states~\cite{Yang2021_Second}, thus leading to the more explicit expression
\be\label{eqapp:corr_scaling_func_high-density}
s^\gamma_{\rm c}[\bxi,n]=\dfrac{E^{\bxi,\gamma}_{\rm c}[n]}{E^{\gamma}_{\rm c}[n]}=s_{\rm c}[\bxi,n_{1/\gamma}]\underset{\gamma\rightarrow 0}{\rightarrow} \dfrac{E^{\bxi,{\rm GL2}}_{\rm c}[n]}{E^{{\rm GL2}}_{\rm c}[n]}, 
\ee
where $E^{{\rm GL2}}_{\rm c}[n]$ is the regular ground-state GL2 correlation functional. Eq.~(\ref{eqapp:corr_scaling_func_high-density}) clearly shows the non-trivial density dependence of the correlation scaling functional.\\

If we now consider the low-density limit instead (\ie, $\gamma\rightarrow +\infty$ in $n_{1/\gamma}$), by analogy with Ref.~\citenum{Gould2023_Electronic}, it comes from Eqs.~(\ref{eqapp:exact_scaling_relation_Fxi}) and (\ref{eqapp:exact_scaling_relation_Tsxi})  
\be\label{eqapp:low-dens-limit_Nc_ens_Hxc_func}
\begin{split}
E^{\bxi}_{\rm Hxc}[n_{1/\gamma}]
&=F^{\bxi}[n_{1/\gamma}]- T^{\bxi}_{\rm s}[n_{1/\gamma}]
=\dfrac{1}{\gamma^2}\left(F^{\bxi,\gamma}[n]-T^{\bxi}_{\rm s}[n]\right)
\\
&\underset{\gamma\rightarrow \infty}{\sim} \dfrac{1}{\gamma^2}\left(\gamma W^{\bxi,{\rm SCE}}_{\rm ee}[n]- T^{\bxi}_{\rm s}[n]\right)
\underset{\gamma\rightarrow \infty}{\sim}\dfrac{1}{\gamma}W^{\bxi,{\rm SCE}}_{\rm ee}[n],  
\end{split}
\ee
where $W^{\bxi,{\rm SCE}}_{\rm ee}[n]$ is the extension to $N$-centered ensembles of the Levy--Lieb functional for TGOK ensembles of strictly correlated electrons (SCE)~\cite{Gould2023_Electronic}, so that, with the justified simplification of the exchange scaling functional in Eq.~(\ref{eqapp:simplification_sx_no_dens_dep}), 
\be
s_{\rm Hx}(\bxi)E_{\rm Hx}[n_{1/\gamma}]+s_{\rm c}[\bxi,n_{1/\gamma}]E_{\rm c}[n_{1/\gamma}]\underset{\gamma\rightarrow \infty}{\sim}\dfrac{1}{\gamma}W^{\bxi,{\rm SCE}}_{\rm ee}[n],
\ee
or, equivalently,
\be
\left(s_{\rm Hx}(\bxi)-s_{\rm c}[\bxi,n_{1/\gamma}]\right)E_{\rm Hx}[n_{1/\gamma}]+s_{\rm c}[\bxi,n_{1/\gamma}]\dfrac{1}{\gamma}W^{{\rm SCE}}_{\rm ee}[n]\underset{\gamma\rightarrow \infty}{\sim}\dfrac{1}{\gamma}W^{\bxi,{\rm SCE}}_{\rm ee}[n],
\ee
where we applied Eq.~(\ref{eqapp:low-dens-limit_Nc_ens_Hxc_func}) in the $\bxi=0$  limit (for which the regular ground-state SCE functional $W^{{\rm SCE}}_{\rm ee}[n]=W^{\bxi=0,{\rm SCE}}_{\rm ee}[n]$ is recovered).   
Therefore, according to Eq.~(\ref{eqapp:exact_scaling_constraint_Nc_ens_Hx}) taken in the $\bxi=0$ limit,
\be
s_{\rm Hx}(\bxi)E_{\rm Hx}[n]+s_{\rm c}[\bxi,n_{1/\gamma}]\left(W^{{\rm SCE}}_{\rm ee}[n]-E_{\rm Hx}[n]\right)\underset{\gamma\rightarrow \infty}{=}W^{\bxi,{\rm SCE}}_{\rm ee}[n],
\ee
or, equivalently,
\be
s_{\rm c}[\bxi,n_{1/\gamma}]\underset{\gamma\rightarrow \infty}{=}\dfrac{W^{\bxi,{\rm SCE}}_{\rm ee}[n]-s_{\rm Hx}(\bxi)E_{\rm Hx}[n]}{W^{{\rm SCE}}_{\rm ee}[n]-E_{\rm Hx}[n]}.
\ee
Finally, by extending the following relation~\cite{Gould2023_Electronic} to $N$-centered ensembles, in analogy with Eq.~(\ref{eqapp:simp_ens_Hx_func_scaled_dens}),
\be
\dfrac{1}{\gamma}W^{\bxi,{\rm SCE}}_{\rm ee}[n]=W^{\bxi,{\rm SCE}}_{\rm ee}[n_{1/\gamma}],
\ee
we obtain the following exact expression of the scaling correlation functional in the low-density limit, 
\be
s_{\rm c}[\bxi,n_{1/\gamma}]\underset{\gamma\rightarrow \infty}{=}\dfrac{W^{\bxi,{\rm SCE}}_{\rm ee}[n_{1/\gamma}]-s_{\rm Hx}(\bxi)E_{\rm Hx}[n_{1/\gamma}]}{W^{{\rm SCE}}_{\rm ee}[n_{1/\gamma}]-E_{\rm Hx}[n_{1/\gamma}]},
\ee
thus confirming its non-trivial density dependence. While, in the context of TGOK ensembles, it was shown that $W^{\bxi,{\rm SCE}}_{\rm ee}[n]=W^{{\rm SCE}}_{\rm ee}[n]$~\cite{Gould2023_Electronic}, 
the evaluation of the strictly correlated functional in the more general context of $N$-centered ensembles, where states with different electron numbers are mixed, is expected to be more subtle, as shown in 1D ~\cite{mirtschink2013derivative}, for example.\\  

In conclusion, the exact uniform coordinate scaling relations fulfilled by the $N$-centered ensemble Hxc functional lead to the following scaled EDFA,
\be\label{eqapp:scaled_EDFA_final_form}
E^{\bxi}_{\rm Hxc}[n]\approx s_{\rm Hx}(\bxi)E_{\rm Hx}[n]+s_{\rm c}[\bxi,n]E_{\rm c}[n],
\ee
and, most importantly, for the computation of Fukui functions, to the following ensemble density-functional Hxc potential approximation,
\be\label{eqapp:scaled_EDFA_Hxc_pot_final_form}
v^{\bxi}_{\rm Hxc}[n](\br)\approx s_{\rm Hx}(\bxi)v_{\rm Hx}[n](\br)+s_{\rm c}[\bxi,n]v_{\rm c}[n](\br)+E_{\rm c}[n]\dfrac{\delta s_{\rm c}[\bxi,n]}{\delta n(\br)},
\ee
where the ensemble weight dependence is incorporated into both the (to-be-determined) Hx and correlation scaling functionals. Interestingly, the density dependence of the latter functional gives additional flexibility when it comes to reproduce any Fukui function pointwise in real space (as shown more explicitly in the upcoming Eq.~(\ref{eqapp:extraction_sc_from_Fukui_functions})). We note at this point that any regular semi-local DFA is now {\it made weight-dependent} thanks to the scaling functionals and, therefore, it will exhibit the expected derivative discontinuity in the Hxc kernel, by construction, with no need to turn to the  more involved hybrid DFT methods.

\subsection{Learning the Hx and correlation scaling functionals in the zero-weight limit}\label{sec:learning_in_zero_weight_limit}

We will assume for now that we have some reference ionization potentials (IPs) and Fukui functions that we would like to reproduce with the scaled EDFA described previously (a self-contained learning procedure will be described later in Sec.~\ref{sec:self-contained_learning}). These data can be generated by a (range-separated) hybrid DFT calculation, for example, or highly accurate quantum chemistry methods, if possible. Our goal is to show that a semi-local DFA, combined with properly adjusted (from the exact $N$-centered ensemble density-functional expressions discussed in the Letter) weight-dependent scaling functionals, can reproduce these data in the regular $N$-electron ground-state ($\bxi=0$) limit of the theory. The learning process is detailed in the following.\\

In practice, it is convenient to use two different $N$-centered ensembles for dealing with the electron addition ($\bxi\equiv \xi_+$ in this case) and removal ($\bxi\equiv \xi_-$ in this case) separately. On that basis, the exact IPs theorem of $N$-centered EDFT reads as follows in the zero-weight limit of the theory (see Eq.~(40) of Ref.~\citenum{dupuy2026ensembledensityfunctionaltheory}):
\be
\begin{split}
I^{N+\frac{1}{2}\pm \frac{1}{2}}=
&=\pm\left(E_0^N-E_0^{N\pm 1}\right)
\\
&=-\varepsilon_{N+\frac{1}{2}\pm \frac{1}{2}}-\dfrac{\left(E_{\rm Hxc}[n_0]-(v_{\rm Hxc}\vert n_0)\right)}{N}\mp\left.\dfrac{\partial{E^{\xi_{\pm}}_{\rm Hxc}[n_0]}}{\partial \xi_\pm}\right|_{\xi_\pm=0}.
\end{split}
\ee
By analogy with optimally-tuned range-separated hybrid functionals, we can start by adjusting the Hx scaling function only, such that the above theorem is fulfilled, which reads explicitly
\be
\begin{split}
I^{N+\frac{1}{2}\pm \frac{1}{2}}\overset{!}{=}
-\varepsilon_{N+\frac{1}{2}\pm \frac{1}{2}}-\dfrac{\left(E_{\rm Hxc}[n_0]-(v_{\rm Hxc}\vert n_0)\right)}{N}\mp E_{\rm Hx}[n_0]\left.\dfrac{\partial{s_{\rm Hx}(\xi_{\pm})}}{\partial \xi_\pm}\right|_{\xi_\pm=0},
\end{split}
\ee
or, equivalently,
\be\label{eqapp:dsHx_over_dxi_zero_weight_from_IP_theo}
\pm\left.\dfrac{\partial{s_{\rm Hx}(\xi_{\pm})}}{\partial \xi_\pm}\right|_{\xi_\pm=0}
=-\dfrac{I^{N+\frac{1}{2}\pm \frac{1}{2}}+\varepsilon_{N+\frac{1}{2}\pm \frac{1}{2}}+\dfrac{\left(E_{\rm Hxc}[n_0]-(v_{\rm Hxc}\vert n_0)\right)}{N}}{E_{\rm Hx}[n_0]}.
\ee
Then, from the exact expression of the Fukui functions derived in the Letter, 
\begin{equation} 
    \begin{split}
        f_{\pm} &=   \big(1+\chi \star f_{\rm Hxc}\big) \star f_{s,\pm} -\; \chi \star f_{\rm Hxc} \star \frac{n_0}{N}
        \\
        &\quad  \pm \; \chi \star \left(\frac{\partial v_{\rm Hxc}^{\bxi}[n]}{\partial \xi_{\pm}}\Big|_{n=n_0}^{\bxi=0}\right), 
    \end{split}
\end{equation}
we can learn about the scaling correlation functional as follows,
\begin{equation} 
    \begin{split}
        f_{\pm} &\overset{!}{=}   \big(1+\chi \star f_{\rm Hxc}\big) \star f_{s,\pm} -\; \chi \star f_{\rm Hxc} \star \frac{n_0}{N}
        \\
        &\quad  \pm \left.\dfrac{\partial{s_{\rm Hx}(\xi_{\pm})}}{\partial \xi_\pm}\right|_{\xi_\pm=0}\;\chi \star v_{\rm Hx}[n_0]
        \\
        &\quad 
        \pm \left.\dfrac{\partial{s_{\rm c}[\xi_{\pm},n_0]}}{\partial \xi_\pm}\right|_{\xi_\pm=0}\;\chi \star v_{\rm c}[n_0]
      \\
&\quad  
\pm E_{\rm c}[n_0]\;\chi\star \left.\dfrac{\delta }{\delta n}\left(\left.\dfrac{\partial s_{\rm c}[\xi_{\pm},n]}{\partial \xi_\pm}\right|_{\xi_\pm=0}\right)\right|_{n=n_0}, 
\end{split}
\end{equation}
which reads more explicitly at any position $\br$ in space, 
\be
\begin{split}
&\pm\left.\dfrac{\partial{s_{\rm c}[\xi_{\pm},n_0]}}{\partial \xi_\pm}\right|_{\xi_\pm=0}v_{\rm c}[n_0](\br)
\pm E_{\rm c}[n_0]\left.\dfrac{\delta }{\delta n(\br )}\left(\left.\dfrac{\partial s_{\rm c}[\xi_{\pm},n]}{\partial \xi_\pm}\right|_{\xi_\pm=0}\right)\right|_{n=n_0}
\\
&\overset{!}{=}\chi^{-1}f_{\pm}(\br)-\chi_s^{-1}f_{s,\pm}(\br)+\dfrac{1}{N}f_{\rm Hxc}n_0(\br)\mp \left.\dfrac{\partial{s_{\rm Hx}(\xi_{\pm})}}{\partial \xi_\pm}\right|_{\xi_\pm=0}\;v_{\rm Hx}[n_0](\br),
\end{split}
\ee
thus leading to the final expression, according to the Dyson equation,
\be\label{eqapp:extraction_sc_from_Fukui_functions}
\begin{split}
&\pm\left.\dfrac{\partial{s_{\rm c}[\xi_{\pm},n_0]}}{\partial \xi_\pm}\right|_{\xi_\pm=0}v_{\rm c}[n_0](\br)
\pm E_{\rm c}[n_0]\left.\dfrac{\delta }{\delta n(\br )}\left(\left.\dfrac{\partial s_{\rm c}[\xi_{\pm},n]}{\partial \xi_\pm}\right|_{\xi_\pm=0}\right)\right|_{n=n_0}
\\
&\overset{!}{=}\chi_s^{-1}\left(f_{\pm}(\br)-f_{s,\pm}(\br)\right)-f_{\rm Hxc}\left(f_{\pm}(\br)-\dfrac{1}{N}n_0(\br)\right)\mp \left.\dfrac{\partial{s_{\rm Hx}(\xi_{\pm})}}{\partial \xi_\pm}\right|_{\xi_\pm=0}v_{\rm Hx}[n_0](\br).
\end{split}
\ee
In order to implement the above (pointwise) constraint, we may first estimate the {\it position-independent} weight derivative $\left.{\partial{s_{\rm c}[\xi_{\pm},n_0]}}/{\partial \xi_\pm}\right|_{\xi_\pm=0}$ contribution, for example, by neglecting its position-dependent density-functional derivative counterpart (the second term on the left-hand side of Eq.~(\ref{eqapp:extraction_sc_from_Fukui_functions})) and integrating over space the correlation potential weighted by the density, \ie,
\be\label{eqapp:scaled_corr_weight_deriv_only_integration}
\begin{split}
&\pm\left.\dfrac{\partial{s_{\rm c}[\xi_{\pm},n_0]}}{\partial \xi_\pm}\right|_{\xi_\pm=0}
:=\dfrac{1}{\int d\br\; v_{\rm c}[n_0](\br)n_0(\br)}\int d\br\; n_0(\br)
\\
&\times
\left[\chi_s^{-1}\left(f_{\pm}(\br)-f_{s,\pm}(\br)\right)-f_{\rm Hxc}\left(f_{\pm}(\br)-\dfrac{1}{N}n_0(\br)\right)\mp \left.\dfrac{\partial{s_{\rm Hx}(\xi_{\pm})}}{\partial \xi_\pm}\right|_{\xi_\pm=0} v_{\rm Hx}[n_0](\br)\right]
,
\end{split}
\ee
and then extract $\left.{\delta }(\left.{\partial s_{\rm c}[\xi_{\pm},n]}/{\partial \xi_\pm}\right|_{\xi_\pm=0})/{\delta n(\br )}\right|_{n=n_0}$, at each position $\br$, from Eq.~(\ref{eqapp:extraction_sc_from_Fukui_functions}). The latter density-functional derivative can actually be re-incorporated into Eq.~(\ref{eqapp:scaled_corr_weight_deriv_only_integration}), thus giving an updated value for $\left.{\partial{s_{\rm c}[\xi_{\pm},n_0]}}/{\partial \xi_\pm}\right|_{\xi_\pm=0}$. This strategy provides an iterative implementation of Eq.~(\ref{eqapp:extraction_sc_from_Fukui_functions}), for a given value of $\left.{\partial{s_{\rm Hx}(\xi_{\pm})}}/{\partial \xi_\pm}\right|_{\xi_\pm=0}$, which has been determined from the IPs theorem, according to Eq.~(\ref{eqapp:dsHx_over_dxi_zero_weight_from_IP_theo}). The full procedure becomes iterative with respect to the latter Hx derivative too if the correlation scaling functional is re-introduced into the expression of the IPs, \ie,
\be
\begin{split}
I^{N+\frac{1}{2}\pm \frac{1}{2}}&\overset{!}{=}
-\varepsilon_{N+\frac{1}{2}\pm \frac{1}{2}}-\dfrac{\left(E_{\rm Hxc}[n_0]-(v_{\rm Hxc}\vert n_0)\right)}{N}\mp E_{\rm Hx}[n_0]\left.\dfrac{\partial{s_{\rm Hx}(\xi_{\pm})}}{\partial \xi_\pm}\right|_{\xi_\pm=0}
\\
&\quad\mp E_{\rm c}[n_0]\left.\dfrac{\partial{s_{\rm c}[\xi_{\pm},n_0]}}{\partial \xi_\pm}\right|_{\xi_\pm=0},
\end{split}
\ee
thus leading to the updated $\left.{\partial{s_{\rm Hx}(\xi_{\pm})}}/{\partial \xi_\pm}\right|_{\xi_\pm=0}$ value:
\be\label{eqapp:taking_sc_into_account_IP_zero_weight_limit}
\pm\left.\dfrac{\partial{s_{\rm Hx}(\xi_{\pm})}}{\partial \xi_\pm}\right|_{\xi_\pm=0}
\overset{!}{=}-\dfrac{I^{N+\frac{1}{2}\pm \frac{1}{2}}+\varepsilon_{N+\frac{1}{2}\pm \frac{1}{2}}+\dfrac{\left(E_{\rm Hxc}[n_0]-(v_{\rm Hxc}\vert n_0)\right)}{N}
\pm E_{\rm c}[n_0]\left.\dfrac{\partial{s_{\rm c}[\xi_{\pm},n_0]}}{\partial \xi_\pm}\right|_{\xi_\pm=0}
}{E_{\rm Hx}[n_0]}.
\ee

\subsection{Self-contained learning of the scaling functionals}\label{sec:self-contained_learning}

The previous construction of Hx and correlation scaling density functionals aimed at reproducing with a semi-local EDFA based on the ansatz of Eq.~(\ref{eqapp:scaled_EDFA_final_form}) any desired IPs and Fukui functions. If such data are not available, thus preventing a direct learning in the zero-weight limit of the theory, the present formalism still allows for the construction of ensemble scaling DFAs from scratch (\ie, with no {\it a priori} knowledge about the targeted Fukui functions), thus making the learning process fully self-contained. For that purpose, we exploit exact properties of the theory, namely the fact that both the IPs and the Fukui functions can be determined, in principle exactly, either by linear interpolation or differentiation of the $N$-centered ensemble energy and density, respectively. This reads more explicitly as follows (see Eq.~(40) of Ref.~\citenum{dupuy2026ensembledensityfunctionaltheory}): 
\begin{subequations}
\begin{align}
I^{N+\frac{1}{2}\pm \frac{1}{2}}
&\underset{\forall \xi_\pm}{=}-\varepsilon^{\xi_\pm}_{N+\frac{1}{2}\pm \frac{1}{2}}-\dfrac{\left(E^{\xi_\pm}_{\rm Hxc}[n^{\xi_\pm}]-(v^{\xi_\pm}_{\rm Hxc}\vert n^{\xi_\pm})\right)}{N}+\left(\dfrac{\xi_\pm}{N}\mp 1\right)\left.\dfrac{\partial{E^{\xi_{\pm}}_{\rm Hxc}[n]}}{\partial \xi_\pm}\right|_{n=n^{\xi_\pm}}
\\
\label{eqapp:IPs_from_LIM}
&=\pm\left(E^{\xi_\pm=0}-\frac{N\pm 1}{N}E^{\xi_\pm=\frac{N}{N\pm 1}}\right),
\end{align}
\end{subequations}
and, as demonstrated in the Letter,
\begin{subequations}
    \begin{align}
        f_\pm 
        &\underset{\forall \xi_\pm}{=}  \big(1+\chi^{\xi_\pm} \star f_{\rm Hxc}^{\xi_\pm}\big) \star f^{\xi_\pm}_{s,\pm} - \chi^{\xi_\pm} \star f_{\rm Hxc}^{\xi_\pm} \star \frac{n^{\xi_\pm}}{N} 
        -\left(\dfrac{\xi_\pm}{N}\mp 1 \right) \chi^{\xi_\pm} \star \frac{\partial v_{\rm Hxc}^{\xi_\pm}[n]}{\partial \xi_{\pm}}\Big|_{n=n^{\xi_\pm}}
\\
\label{eqapp:Fukui_funcs_from_LIM}
&=\pm\left(\frac{N\pm 1}{N}n^{\xi_\pm=\frac{N}{N\pm 1}}-n^{\xi_\pm=0}\right),
\end{align}
\end{subequations} 
or, equivalently, 
\be\label{eqapp:exact_key_eq_IPs_learning_weight_dep_from_scratch}
\left.\dfrac{\partial{E^{\xi_{\pm}}_{\rm Hxc}[n]}}{\partial \xi_\pm}\right|_{n=n^{\xi_\pm}}\underset{\forall \xi_\pm}{=} \dfrac{\pm\left(E^{\xi_\pm=0}-\frac{N\pm 1}{N}E^{\xi_\pm=\frac{N}{N\pm 1}}\right)+\varepsilon^{\xi_\pm}_{N+\frac{1}{2}\pm \frac{1}{2}}
+\dfrac{\left(E^{\xi_\pm}_{\rm Hxc}[n^{\xi_\pm}]-(v^{\xi_\pm}_{\rm Hxc}\vert n^{\xi_\pm})\right)}{N}
}{\left(\dfrac{\xi_\pm}{N}\mp 1\right)},
\ee
and, according to the ensemble Dyson equation,
\be\label{eqapp:exact_key_eq_Fukuifuncs_learning_weight_dep_from_scratch}
\begin{split}
\frac{\partial v_{\rm Hxc}^{\xi_\pm}[n]}{\partial \xi_{\pm}}\Big|_{n=n^{\xi_\pm}}
&\underset{\forall \xi_\pm}{=}-\dfrac{
 \left[\chi_s^{\xi_\pm}\right]^{-1}\left(\pm\left(\frac{N\pm 1}{N}n^{\xi_\pm=\frac{N}{N\pm 1}}-n^{\xi_\pm=0}\right)-f^{\xi_\pm}_{s,\pm}\right)}{\left(\dfrac{\xi_\pm}{N}\mp 1 \right)} 
\\
&\quad
+\dfrac{f_{\rm Hxc}^{\xi_\pm}\left(\pm\left(\frac{N\pm 1}{N}n^{\xi_\pm=\frac{N}{N\pm 1}}-n^{\xi_\pm=0}\right)-\frac{n^{\xi_\pm}}{N}\right)}{\left(\dfrac{\xi_\pm}{N}\mp 1 \right)}.
\end{split}
\ee
Eqs.~(\ref{eqapp:exact_key_eq_IPs_learning_weight_dep_from_scratch}) and (\ref{eqapp:exact_key_eq_Fukuifuncs_learning_weight_dep_from_scratch}) are the key in-principle exact relations from which we can learn self-consistently about the weight dependence of the Hx and correlation scaling functionals over the full range of weight values $0\leq \xi_\pm\leq \frac{N}{N\pm 1}$. While, in the initial reference calculation from which the to-be-learned (along the lines of the previous Sec.~\ref{sec:learning_in_zero_weight_limit}) IPs and Fukui functions  will be evaluated (according to Eqs.~(\ref{eqapp:IPs_from_LIM}) and (\ref{eqapp:Fukui_funcs_from_LIM}), respectively), a regular (weight-independent) semi-local DFA can be employed, which consists in making the following approximations,
\be
s_{\rm Hx}(\xi_\pm)\approx 1 \approx s_{\rm c}[\xi_\pm,n],
\ee
the knowledge of the scaling functionals that we will gain, ultimately, in the limiting $\xi_\pm=\frac{N}{N\pm 1}$ cases, will provide updated reference values (the learned scaling functionals allowing for their re-evaluation, this time at the scaled KS-EDFA level of calculation), thus initiating the self-consistency of what we refer to as a self-contained learning. We detail in the following the learning process, which generalizes that of Sec.~\ref{sec:learning_in_zero_weight_limit} to all weight values.\\

We start with the analog of Eq.~(\ref{eqapp:dsHx_over_dxi_zero_weight_from_IP_theo}) for any weight value $\xi_\pm$, which is deduced from Eq.~(\ref{eqapp:exact_key_eq_IPs_learning_weight_dep_from_scratch}),   
\be\label{eqapp:Hx_scaling_func_only_from_IP}
\dfrac{\partial s_{\rm Hx}(\xi_\pm)}{\partial \xi_\pm}\approx \dfrac{\pm\left(E^{\xi_\pm=0}-\frac{N\pm 1}{N}E^{\xi_\pm=\frac{N}{N\pm 1}}\right)+\varepsilon^{\xi_\pm}_{N+\frac{1}{2}\pm \frac{1}{2}}
+\dfrac{\left(E^{\xi_\pm}_{\rm Hxc}[n^{\xi_\pm}]-(v^{\xi_\pm}_{\rm Hxc}\vert n^{\xi_\pm})\right)}{N}
}{\left(\dfrac{\xi_\pm}{N}\mp 1\right)E_{\rm Hx}[n^{\xi_\pm}]},
\ee
and the analog of Eq.~(\ref{eqapp:scaled_corr_weight_deriv_only_integration}), which is deduced from Eq.~(\ref{eqapp:exact_key_eq_Fukuifuncs_learning_weight_dep_from_scratch}), 
\be\label{eqapp:extract_deriv_sc_anyweight_from_average_Fukui_functions}
\begin{split}
\frac{\partial s_{\rm c}[{\xi_\pm},n]}{\partial \xi_{\pm}}\Big|_{n=n^{\xi_\pm}}
&
\approx \dfrac{1}{\left(\dfrac{\xi_\pm}{N}\mp 1 \right)\int d\br\, n^{\xi_\pm}(\br)v_{\rm c}[n^{\xi_\pm}](\br)} \int d\br\, n^{\xi_\pm}(\br)
\\
&\times \Bigg(-\left[\chi_s^{\xi_\pm}\right]^{-1}\left(\pm\left(\frac{N\pm 1}{N}n^{\xi_\pm=\frac{N}{N\pm 1}}-n^{\xi_\pm=0}\right)-f^{\xi_\pm}_{s,\pm}\right)
\\
&\quad\quad 
+f_{\rm Hxc}^{\xi_\pm}\left(\pm\left(\frac{N\pm 1}{N}n^{\xi_\pm=\frac{N}{N\pm 1}}-n^{\xi_\pm=0}\right)-\frac{n^{\xi_\pm}}{N}\right) 
\\
&\quad \quad 
-\left(\dfrac{\xi_\pm}{N}\mp 1 \right)\dfrac{\partial s_{\rm Hx}(\xi_\pm)}{\partial \xi_\pm}v_{\rm Hx}[n^{\xi_\pm}]\Bigg)(\br),
\end{split}
\ee
where the density-functional derivative of the correlation scaling functional has been neglected, as a convenient practical approximation. If necessary, the latter approximation can be improved, by extending the pointwise Eq.~(\ref{eqapp:extraction_sc_from_Fukui_functions}) to any weight value, from the in-principle exact Eq.~(\ref{eqapp:exact_key_eq_Fukuifuncs_learning_weight_dep_from_scratch}) and the ansatz of Eq.~(\ref{eqapp:scaled_EDFA_Hxc_pot_final_form}).
By analogy with Eq.~(\ref{eqapp:taking_sc_into_account_IP_zero_weight_limit}), the value of the Hx scaling function weight derivative deduced from Eq.~(\ref{eqapp:Hx_scaling_func_only_from_IP}) can be further improved, in principle, if the scaling correlation functional is included into the IPs expression, thus leading to the complete expression
\be\label{eqapp:improved_sHx_IP_with_sc_any_weight}
\begin{split}
\dfrac{\partial s_{\rm Hx}(\xi_\pm)}{\partial \xi_\pm}&\approx \dfrac{\pm\left(E^{\xi_\pm=0}-\frac{N\pm 1}{N}E^{\xi_\pm=\frac{N}{N\pm 1}}\right)+\varepsilon^{\xi_\pm}_{N+\frac{1}{2}\pm \frac{1}{2}}
+\dfrac{\left(E^{\xi_\pm}_{\rm Hxc}[n^{\xi_\pm}]-(v^{\xi_\pm}_{\rm Hxc}\vert n^{\xi_\pm})\right)}{N}
}{\left(\dfrac{\xi_\pm}{N}\mp 1\right)E_{\rm Hx}[n^{\xi_\pm}]},
\\
&\quad \quad
-\dfrac{E_{\rm c}[n^{\xi_\pm}]}{E_{\rm Hx}[n^{\xi_\pm}]}\frac{\partial s_{\rm c}[{\xi_\pm},n]}{\partial \xi_{\pm}}\Big|_{n=n^{\xi_\pm}},
\end{split}
\ee
which can then be inserted into Eq.~(\ref{eqapp:extract_deriv_sc_anyweight_from_average_Fukui_functions}), so that the correlation scaling functional can be updated consistently.\\

Once the weight derivatives have been evaluated at $\xi_\pm$ (the first calculation being performed for $\xi_\pm=0$, with $s_{\rm Hx}(\xi_\pm=0)=s_{\rm c}[\xi_\pm=0,n]=1$), the ensemble KS calculation can be run for the next weight value $\xi_\pm+\eta$ ($\eta >0$ being a small weight step), with the following ensemble scaled EDFA,
\be\label{eqapp:scaled_Hxc_func_plus_eta}
E^{\xi_\pm+\eta}_{\rm Hxc}[n]&\approx s_{\rm Hx}(\xi_\pm+\eta)E_{\rm Hx}[n]
+s_{\rm c}[\xi_\pm+\eta,n^{\xi_\pm+\eta}]E_{\rm c}[n],
\ee
where 
\be
s_{\rm Hx}(\xi_\pm+\eta)\approx s_{\rm Hx}(\xi_\pm)+\eta\dfrac{\partial s_{\rm Hx}(\xi_\pm)}{\partial \xi_\pm} 
\ee
and
\be
s_{\rm c}[\xi_\pm+\eta,n^{\xi_\pm+\eta}]\approx s_{\rm c}[\xi_\pm,n^{\xi_\pm}]+\eta\left.\dfrac{\partial s_{\rm c}[\xi_\pm,n]}{\partial \xi_\pm}\right|_{n=n^{\xi_\pm}}.
\ee
When reaching the limiting cases $\xi_\pm=\frac{N}{N\pm 1}$, the reference IPs and Fukui functions can be re-evaluated from the scaled EDFA of Eq.~(\ref{eqapp:scaled_Hxc_func_plus_eta}) [$\xi_\pm=\frac{N}{N\pm 1}$ and $\eta=0$ in this case] by linear interpolation, according to Eqs.~(\ref{eqapp:IPs_from_LIM}) and (\ref{eqapp:Fukui_funcs_from_LIM}). The Hx and correlation scaling functionals can then be updated accordingly, by going through the steps described in Eqs.~(\ref{eqapp:Hx_scaling_func_only_from_IP})--(\ref{eqapp:improved_sHx_IP_with_sc_any_weight}), thus making the full procedure self-consistent. A drastic simplification, which avoids evaluating the ensemble linear response function, consists in scaling the Hx functional only (\ie, in assuming that $s_{\rm c}[\xi_\pm,n]\approx 1$). In this case, the learning process relies exclusively on the IPs theorem, as depicted in Eq.~(\ref{eqapp:Hx_scaling_func_only_from_IP}). This formulation, where the strict linear variation in weight of the $N$-centered ensemble density is not guaranteed anymore (because it is not enforced through the correlation scaling functional) and should therefore be scrutinized~\cite{Morgante2023_Strategies}, is expected to be the most practical one. Unlike in range-separated hybrid DFT, where the tuning of the range-separation parameters (which is the only degree of ``freedom'' the theory can offer) echoes the adjustment of the weight-dependent Hx scaling function, the present ensemble scaling strategy offers a clearer path to improvement, through the scaling of the correlation functional.

\section{Analytical expressions for the Hubbard dimer}

\subsection{Exact expressions}\label{sec:exact_Hubbard}

{\bf Model} -- We consider a model diatomic system built from the Hubbard dimer, with Hamiltonian
\begin{equation}
    \hat{H}_e = -t \sum_{\sigma} \big( \hat{a}_{0\sigma}^{\dagger} \hat{a}_{1\sigma} + \hat{a}_{1\sigma}^{\dagger} \hat{a}_{0\sigma} \big) + U \sum_{i=0}^1  \hat{n}_{i \uparrow} \hat{n}_{i \downarrow} + \frac{\Delta v}{2} \big( \hat{n}_1 - \hat{n}_0 \big)
\end{equation}
where operators are written in second quantization and the labels 0 and 1 refer to the first and second atomic sites, respectively. The variable $\sigma=\uparrow,\downarrow$ denotes the spin. We have $\hat{n}_{i \sigma}=\hat{a}_{i\sigma}^{\dagger} \hat{a}_{i\sigma}$ and the total density on site $i$ $\displaystyle \hat{n}_i = \sum_{\sigma=\uparrow \downarrow} \hat{n}_{i \sigma}$. In the (diabatic) $N=2$ electrons singlet basis $\Phi_1=\ket{1_{\uparrow}1_{\downarrow}}$, $\Phi_2=\frac{1}{\sqrt{2}}(\ket{1_{\uparrow}2_{\downarrow}}-\ket{1_{\downarrow}2_{\uparrow}})$, $\Phi_3=\ket{2_{\uparrow}2_{\downarrow}}$, its representation becomes
\begin{equation}
    H_N = \begin{bmatrix}
U-\Delta v & -\sqrt{2} \, t & 0
\\
-\sqrt{2} \, t & 0 & -\sqrt{2} \, t 
\\
0 & -\sqrt{2} \, t & U+\Delta v
\end{bmatrix}
\end{equation}
the exact energies being analytically given by the third-order polynomial equation\cite{carrascal2015hubbard,carrascal2016corrigendum,deur2017exact}:
\begin{equation} \label{eqapp:polynomial}
    E_{N,i}^3 = -4 t^2 U + (4 t^2 - U^2 + \Delta v^2) E_{N,i} + 2U \, E_{N,i}^2
\end{equation}
In particular, the $N=2$-electron ground state energy, that is the lowest solution to Eq.~\eqref{eqapp:polynomial}, reads\cite{carrascal2015hubbard,carrascal2016corrigendum}
\begin{equation} \label{eqapp:En}
    E_N(\Delta v) = \frac{4t}{3} \bigg(u-w \sin \left(\theta + \frac{\pi}{6}\right) \bigg)
\end{equation}
The angle $\theta$ is defined as
\begin{equation} \label{eqapp:costheta}
    \cos (3 \theta) = \frac{u}{w^3} \Big( 9 \big(\nu^2-1/2 \big) -u^2 \Big)
\end{equation}
with $u=U / 2t$, $w = \sqrt{3(1+\nu^2) + u^2}$ and $\nu = \Delta v / 2t$.

\bigskip 
{\bf Ensemble density and Fukui functions from the external potential} -- The ensemble is comprised of the 1-to-3 electron(s) ground-states, which all separately satisfy the variational principle
\begin{equation}
    E_{N\pm p}[v] = \underset{n_N\pm p}{\rm min} \Big\{ F [n_{N\pm p}] + \int d {\bf r} \; v({\bf r}) \; n_{N\pm p}({\bf r}) \Big\}
\end{equation}
with $F[n_{N\pm p}]$ the universal Hohenberg-Kohn functional. By the means of a Legendre-Fenchel transform, we can obtain an expression for the latter\cite{deur2018exploring}:
\begin{equation}
    F[n_{N\pm p}] = \underset{v}{\rm sup} \Big\{ E_{N\pm p} [v] - \int d {\bf r} \; v({\bf r}) \; n_{N\pm p}({\bf r}) \Big\}
\end{equation}
from which we can deduce the link between density and energy-derivative
\begin{equation}
    n_{N\pm p}({\bf r}) = \frac{\delta E_{N\pm p} [v]}{\delta v({\bf r})}
\end{equation}

In Hubbard dimer, the Site Occupation Functional Theory (SOFT) equivalent is
\begin{equation} \label{eqapp:ind_Liebmax}
    F_{N\pm p}(\Delta n_{N\pm p}) = \underset{\Delta v}{\rm sup} \Big\{ E_{N\pm p} [\Delta v] - \frac{\Delta v}{2} \; \Delta n_{N\pm p} \Big\}
\end{equation}

with $\Delta n_{N\pm p}$ the density difference between site 1 and site 0. As the sum of density on both sites always equals $N\pm p$, we can use the density on site 0, noted $n_{N\pm p}$, as the basic variable
\begin{equation} \label{eqapp:Dn_dedv}
    \frac{\Delta n_{N\pm p}}{2} = \frac{N\pm p}{2} - n_{N\pm p} = \frac{\partial E_{N\pm p} [\Delta v]}{\partial \Delta v}
\end{equation}
To obtain the ground-state energy of the 1-electron system we note it lacks electronic correlation entirely, and is thus given by the lowest orbital energy in the non-interacting 2-electron system\cite{senjean2018unified} with external potential $\Delta v$
\begin{equation} \label{eqapp:Enm}
    E_{N_-} (\Delta v) = - \sqrt{t^2 + (\Delta v/2)^2}
\end{equation}
while hole-particle symmetry\cite{senjean2018unified} allows to obtain the 3-electrons ground-state energy as
\begin{equation} \label{eqapp:Enp}
    E_{N_+} (\Delta v) = U - \sqrt{t^2 + (\Delta v/2)^2}
\end{equation}

Inserting Eqs.~(\ref{eqapp:En},\ref{eqapp:Enm},\ref{eqapp:Enp}) in Eq.~\eqref{eqapp:Dn_dedv} yields
\begin{equation} \label{eqapp:ind_dens}
    \left\{\begin{split}
         \; & n_{N_-} (\Delta v)  = \frac{1}{2} + \frac{1}{4} \frac{\Delta v}{\sqrt{t^2+(\Delta v/2})^2}
        \\
        & n_{N} (\nu) = 1 + \frac{2\nu}{w} \sin \left(\theta + \frac{\pi}{6}\right) + \Theta  \cos \left(\theta + \frac{\pi}{6}\right) 
        \\
        & n_{N_+} (\Delta v) = \frac{3}{2} + \frac{1}{4} \frac{\Delta v}{\sqrt{t^2+(\Delta v/2})^2}
    \end{split}\right.
\end{equation}
where 
\begin{equation}
    \Theta = -\frac{4\nu}{w \sin(3\theta)} \left( \frac{u}{w} -\frac{1}{2} \cos(3 \theta) \right)
\end{equation}

From Eqs.~\eqref{eqapp:ind_dens}, computation of the ensemble density is straightforward
\begin{equation} \label{eqapp:ens_density_pot}
    n^{\bxi}(\Delta v) = \xi_+ \, n_{N_+}(\Delta v) + \xi_- \, n_{N_-}(\Delta v) + \left(1-\frac{3 \xi_+ + \xi_-}{2}\right) n_{N}(\Delta v)
\end{equation}
as is the computation of Fukui functions. The latter are defined as the difference between densities on site 0 for neighboring $N$ and $N\pm1$-electron ground states
\begin{equation}
        \left\{\begin{split}
         \; & f_-  (\Delta v)  = n_{N}(\Delta v) - n_{N_-}(\Delta v)
        \\
        & f_+  (\Delta v) = n_{N_+}(\Delta v) - n_{N}(\Delta v)
    \end{split}\right.
\end{equation}

\bigskip
{\bf KS Fukui functions and KS ensemble response from the KS potential} -- Each KS ground state within the ensemble satisfies the variational principle
\begin{equation} \label{eqapp:KS_var_principle}
    \mathcal{E}_{N\pm p}(\Delta v^{\bxi}_s) = {\rm min}_{\Delta n_{N\pm p}} \left\{ T_s(\Delta n) + \Delta n_{N\pm p} \,  \frac{\Delta v^{\bxi}_{s}}{2} \right\}
\end{equation}
The minimizing, weight-dependent KS density differences are written $\Delta n^{\bxi}_{s,N\pm p}$. Differentiating the above expression with respect to the ensemble KS potential yields
\begin{equation} \label{eqapp:Deltan}
    \frac{\partial \mathcal{E}_{N\pm p}}{\partial \Delta v^{\bxi}_{s}}(\Delta n^{\bxi}_{s,N\pm p}) = \frac{\Delta n^{\bxi}_{s,N\pm p}}{2}
\end{equation}

To obtain the total KS energies, we note KS orbital energies in the Hubbard dimer are 
\begin{equation} \label{eqapp:KS_orbital_ene}
    \begin{split}
        \epsilon_{0,1} (\Delta v^{\bxi}_s) & = \pm \,\epsilon_{\rm H} (\Delta v^{\bxi}_s) 
        \\
        \epsilon_{\rm H} (\Delta v^{\bxi}_s) & = - \sqrt{t^2 + (\Delta v^{\bxi}_s/2)^2}
    \end{split} 
\end{equation}
so that filling the 1- to 3-electron ground states according to the Aufbau principle yields
\begin{equation}
    \left\{
    \begin{split}
        \mathcal{E}_{N_-}(\Delta v^{\bxi}_s) & = - \sqrt{t^2 + (\Delta v^{\bxi}_s/2)^2}
        \\
            \mathcal{E}_{N}(\Delta v^{\bxi}_s) & = - 2 \sqrt{t^2 + (\Delta v^{\bxi}_s/2)^2}
        \\
        \mathcal{E}_{N_+}(\Delta v^{\bxi}_s) & = - \sqrt{t^2 + (\Delta v^{\bxi}_s/2)^2}
    \end{split}
    \right.
\end{equation}

Inserting these expressions in Eq.~\ref{eqapp:Deltan} and using $\Delta n^{\bxi}_{s,N\pm p} = N \pm p - 2\,n^{\bxi}_{s,N\pm p}$ where $n^{\bxi}_{s,N\pm p}$ is KS the density on site 0, one gets individual densities
\begin{equation} \label{eqapp:ind_ks_dens}
    \left\{\begin{split}
         \; & n^{\bxi}_{s,N_-} (\Delta v^{\bxi}_{s})  = \frac{1}{2} + \frac{1}{4} \frac{\Delta v^{\bxi}_{s}}{\sqrt{t^2+(\Delta v^{\bxi}_{s}/2})^2}
        \\
        & n^{\bxi}_{s,N} (\Delta v^{\bxi}_{s}) = 1 + \frac{1}{2} \frac{\Delta v^{\bxi}_{s}}{\sqrt{t^2+(\Delta v^{\bxi}_{s}/2})^2}
        \\
        & n^{\bxi}_{s,N_+} (\Delta v^{\bxi}_{s}) = \frac{3}{2} + \frac{1}{4} \frac{\Delta v^{\bxi}_{s}}{\sqrt{t^2+(\Delta v^{\bxi}_{s}/2})^2}
    \end{split}\right.
\end{equation}

These allows to express the KS Fukui functions as
\begin{equation} \label{eqapp:ks_fukui}
        \left\{\begin{split}
         \; & f^{\bxi}_{s,-}  (\Delta v^{\bxi}_{s})  = \frac{1}{2} + \frac{1}{4} \frac{\Delta v^{\bxi}_{s}}{\sqrt{t^2+(\Delta v^{\bxi}_{s}/2})^2}
        \\
        & f^{\bxi}_{s,+}  (\Delta v^{\bxi}_{s}) = \frac{1}{2} - \frac{1}{4} \frac{\Delta v^{\bxi}_{s}}{\sqrt{t^2+(\Delta v^{\bxi}_{s}/2})^2}
    \end{split}\right.
\end{equation}

To obtain the ensemble KS response function, we first express ensemble density (on site 0) $n^{\bxi}$ as a function of the KS potential. Inserting Eq.~\eqref{eqapp:ind_ks_dens} into the definition
\begin{equation} 
   n^{\bxi} = \xi_+ \, n^{\bxi}_{s,N_+} + \xi_- \, n^{\bxi}_{s,N_-} + \left(1-\frac{3 \xi_+ + \xi_-}{2}\right) n^{\bxi}_{s,N}
\end{equation}
we obtain
\begin{equation} \label{eqapp:ens_dens_dv}
    n^{\bxi}(\Delta v^{\bxi}_{s}) = 1 + \frac{1-\xi_+}{2} \frac{\Delta v^{\bxi}_{s}}{\sqrt{t^2+(\Delta v^{\bxi}_{s}/2})^2}
\end{equation}
Differentiation with respect to the KS potential yields the KS response function, 
\begin{equation} \label{eqapp:ens_ks_resp}
    \chi_s^{\bxi}(\Delta v^{\bxi}_{s}) = \frac{\partial n^{\bxi}(\Delta v_s)}{\partial \Delta v_{s}}\Big|_{\Delta v_s =\Delta v^{\bxi}_{s}} = \frac{(1-\xi_+)t^2}{2 \left(t^2+(\Delta v^{\bxi}_s/2 )^2\right)^{3/2}} 
\end{equation}

\bigskip 
{\bf Ensemble Hxc and KS potentials from the density} --  Inverting Eq.~\eqref{eqapp:ens_dens_dv} yields the ensemble KS potential as a function of ensemble density
\begin{equation} \label{eqapp:Dvs_n}
    \Delta v^{\bxi}_s(n^{\bxi}) = \frac{2 \, t \, (n^{\bxi}-1)}{\sqrt{(1-\xi_+)^2-(n^{\bxi}-1)^2}} 
\end{equation}
while the ensemble Hxc potential is defined as
\begin{equation} 
    v^{\bxi}_{\rm Hxc}[n] := v^{\bxi}_{s}[n] - v^{\bxi}[n]
\end{equation}
where $v^{\bxi}[n]$ is the external potential reproducing the density $n$ for a given set of weights $\bxi$. For the Hubbard Dimer, the SOFT analog is
\begin{equation} \label{eqapp:defSOFTvhxc}
    \Delta v^{\bxi}_{\rm Hxc}(n) = \Delta v^{\bxi}_{s}(n) - \Delta v^{\bxi}(n)
\end{equation}
with $\Delta v^{\bxi}(n)$ being the maximizing potential in the ensemble analog of Eq.~\eqref{eqapp:ind_Liebmax}:
\begin{equation} \label{eqapp:ens_Liebmax}
    F^{\bxi}(n)  = \underset{\Delta v}{\rm sup} \Big\{ \xi_- \, E_{N_-} (\Delta v) + \xi_+ \, E_{N_+} (\Delta v) + \left(1-\frac{3\xi_+ + \xi_-}{2}\right) \, E_{N} (\Delta v) + \Delta v (n-1) \Big\}
\end{equation}
No analytical expression is known from the above equation, but solving the optimization problem for given $n$ and $\bxi$ is straightforward, and can be done to arbitrary accuracy.

\bigskip 
{\bf Ensemble response and Hxc kernel} -- The ensemble response is defined as
\begin{equation}
    \chi^{\bxi}[n^{\bxi}]({\bf r}) := \frac{\delta n^{\bxi}[v] }{\delta v({\bf r})} \Big|_{v=v^{\bxi}}
\end{equation}
with the Hubbard Dimer SOFT analog being
\begin{equation} \label{eqapp:ens_resp_softdef}
    \chi^{\bxi}(n^{\bxi}) = \frac{\partial n^{\bxi}(\Delta v) }{\partial \Delta v} \Big|_{\Delta v=\Delta v^{\bxi}}
\end{equation}

Inserting ensemble density Eq.~\eqref{eqapp:ens_density_pot} (see Eq.~\eqref{eqapp:ind_dens}), we get
\begin{equation} \label{eqapp:ens_resp}
    \chi^{\bxi}(n^{\bxi}) =  \frac{(\xi_+ + \xi_-)}{4} \frac{t^2}{\big[t^2 + (\Delta v/2)^2\big]^{3/2}} - \left(1-\frac{3\xi_+ +\xi_-}{2}\right) E_N''
\end{equation}
where
\begin{equation} 
    E''_N = -\frac{4t}{3}\bigg( \big(w'' - w \, \theta'^2\big) \sin\left(\theta + \frac{\pi}{6}\right) + \big(w \, \theta'' +\frac{3 \nu}{2 w t} \theta'\big) \cos\left(\theta + \frac{\pi}{6}\right) \bigg)
\end{equation}
and
\begin{equation}
\left\{\begin{split}
    \; & w'' = \frac{3}{4t} \frac{3+u^2}{w^3t}
    \\
    & \theta' = - \frac{3 \nu}{w^2 t \sin(3\theta)} \left( \frac{u}{w} - \frac{1}{2} \cos(3\theta) \right)
    \\
    & \theta'' = \frac{3}{2} \frac{1}{w^2t^2 \sin(3\theta)} \left[ \frac{1}{2} \cos(3\theta)-\frac{u}{w} + \frac{3}{2} \frac{\nu^2}{w^2} \left(12 \frac{u}{w} - 5 \cos(3\theta)\right) \right] - 3 \theta'^2 \text{cot}(3\theta)
\end{split} \right.
\end{equation}

Finally, we can compute the Hxc kernel from ensemble Dyson equation
\begin{equation} \label{eqapp:ens_dyson}
    f_{\rm Hxc}^{\bxi}(n^{\bxi}) = \Big( \chi^{\bxi}_s(n^{\bxi}) \Big)^{-1} - \Big( \chi^{\bxi}(n^{\bxi}) \Big)^{-1}
\end{equation}
using Eqs.~\eqref{eqapp:ens_ks_resp}, \eqref{eqapp:Dvs_n} and \eqref{eqapp:ens_resp}.

\subsection{Taylor expansion in the low-correlation limit}\label{sec:Taylor_Hubbard}

In all generality, the ensemble Hxc energy is related to the universal functional and the KS kinetic energy $T_s^{\bxi}[n]$ by
\begin{equation} \label{eqapp:defEhxc}
    E^{\bxi}_{\rm Hxc}[n] := F^{\bxi}[n] - T_s^{\bxi}[n]
\end{equation}
with the SOFT equivalent obtained by replacing functional dependence on a $\bf r$-dependent density by dependence on site occupation. In the low-correlation regime, one can expect a Taylor expansion in $U$ for the universal ensemble functional to be a good approximation:
\begin{equation} \label{eqapp:ens_f_taylor}
    \begin{split}
            F^{\bxi}(n) & := \underset{\Delta v}{\rm sup} \Big\{ \xi_+ \, E_{N_+} (\Delta v) + \xi_- \, E_{N_-} (\Delta v) + \left(1-\frac{3 \xi_+ + \xi_-}{2}\right) E_{N} (\Delta v) + \Delta v (n-1) \Big\}
            \\
            & = \underset{\Delta v}{\rm sup} \Big\{ \xi_{+} U -(\xi_{+}+\xi_{-})  \sqrt{t^2 + (\Delta v/2)^2}  + \left(1-\frac{3 \xi_+ + \xi_-}{2}\right) E_{N} (\Delta v) + \Delta v (n-1) \Big\}
            \\
            & = \sum_{i=0}^{\infty} \frac{U^i}{i!} \frac{\partial^i F^{\bxi}(n)}{\partial U^i}\Big|_{U=0}
    \end{split}
\end{equation}
As the true system becomes identical to its KS analog in the 0-correlation limit,
\begin{equation}
    F^{\bxi}(n) \underset{U\rightarrow 0}{=} T^{\bxi}_s(n)
\end{equation}
we can express the Hxc energy from the Taylor expansion of $F^{\bxi}(n)$:
\begin{equation}
    E^{\bxi}_{\rm Hxc}(n) = \sum_{i=1}^{\infty} \frac{U^i}{i!} \frac{\partial^i F^{\bxi}(n)}{\partial U^i}\Big|_{U=0}
\end{equation}
The first term yields the ensemble Hartree-exchange energy, which we express from Eq.~\eqref{eqapp:ens_f_taylor} as
\begin{equation}
    E^{\bxi}_{\rm Hx}(n) = U \frac{\partial F^{\bxi}(n)}{\partial U}\Big|_{U=0} = U \left[ \xi_{+} + \left(1-\frac{3 \xi_+ + \xi_-}{2}\right) \frac{\partial E_{N} (\Delta v)}{\partial U} \right]_{\Delta v = \Delta v^{\bxi}(n,U=0)}
\end{equation}
where we explicitly indicated the $U$-dependence of $\Delta v^{\bxi}(n)$, that is the external potential reproducing the trial ensemble density $n$ for a given set of weights. Differentiating the polynomial Eq.~\eqref{eqapp:polynomial} with respect to $U$ yields the expression
\begin{equation}
    \frac{\partial E_{N}}{\partial U} \Big|_{U=0} = \frac{\nu^2 + 1/2}{1 + \nu^2}
\end{equation}
From the identity of KS and true systems in the 0-correlation limit,
\begin{equation}
    \nu = \frac{\Delta v^{\bxi}}{2t} \underset{U \rightarrow 0}{=} \frac{\Delta v^{\bxi}_s}{2t}
\end{equation}
we can insert KS potential Eq.~\eqref{eqapp:Dvs_n} and get
\begin{equation} \label{eqapp:ens_ehx}
    E^{\bxi}_{\rm Hx}(n) = \frac{U}{2} \left[ 1 + \frac{\xi_+ - \xi_-}{2} + \left(1-\frac{3 \xi_+ + \xi_-}{2}\right) \frac{(n-1)^2}{(1-\xi_+)^2} \right]
\end{equation}
Using this term alone corresponds to the EEXX approximation in main text.

\rev{A quirk of the model is that the above Hx energy is not of the ansatz form Eq.~\eqref{eqapp:scaled_EDFA_final_form}, but possesses a purely weight-dependent component,}
\rev{\begin{equation}
    E^{\bxi}_{\rm Hx}(n) = s_{\rm Hx}(\bxi) E_{\rm Hx}(n) 
    + \tilde{E}_{\rm Hx}(\bxi)
\end{equation}}
\rev{with}
\rev{\begin{equation}
    s_{\rm Hx}(\bxi) = \frac{1-(\xi_-+3\xi_+)/2}{(1-\xi_+)^2}
\end{equation}}
\rev{and}
\rev{\begin{equation}
    \tilde{E}_{\rm Hx}(\bxi) = \frac{U}{2} \left(\frac{\xi_+-\xi_-}{2} + \frac{\xi_+^2+\xi_-/2-\xi_+/2}{(1-\xi_+)^2}\right)
\end{equation}}
\rev{which does not respect the uniform coordinate scaling relation Eq.~\eqref{eqapp:exact_scaling_constraint_Nc_ens_Hx}. While these relations do not apply directly to SOFT, we note that upon taking $\xi_+=0$, one recovers the ansatz form Eq.~\eqref{eqapp:scaled_EDFA_final_form} and exact scaling relation Eq.~\eqref{eqapp:exact_scaling_constraint_Nc_ens_Hx} for any $\xi_-$. This is linked to the fact the one-electron ground state $|\Psi_+\rangle$ has no electronic repulsion and is thus unaffected by its $\gamma$-scaling.}
\rev{To circumvent this peculiarity of our model system, we will derive scaling functions from the ratios of Hx and correlation potentials, which satisfy the general ansatz (see Sec.~\ref{sec:comp_details}).}

The second order term in the Taylor expansion, which constitutes the PT2 approximation to the correlation energy, reads
\begin{equation} \label{eqapp:d2fdu2}
    \begin{split}
            \frac{\partial^2 F^{\bxi}(n)}{\partial U^2}\Big|_{U=0} =  \left(1-\frac{3 \xi_+ + \xi_-}{2}\right) \bigg[ & \frac{\partial^2 E_{N}(\Delta v)}{\partial U^2}\Big|_{\Delta v = \Delta v^{\bxi}(U=0)} 
            \\
            & + \frac{\partial^2 E_{N}(\Delta v)}{\partial U \partial \Delta v}\Big|_{\Delta v = \Delta v^{\bxi}(U=0)} \frac{\partial \Delta v^{\bxi}(U)}{\partial U}\Big|_{U=0} \bigg]
    \end{split}
\end{equation}

Differentiating the polynomial Eq.~\eqref{eqapp:polynomial} appropriately yields
\begin{equation} \label{eqapp:d2edu2}
    \frac{\partial^2 E_{N}(\Delta v)}{\partial U^2}\Big|_{\Delta v = \Delta v^{\bxi}(U=0)} =  - \frac{1}{2t} \frac{\nu^2+1/4}{(1+\nu^2)^{5/2}}
\end{equation}
and
\begin{equation} \label{eqapp:dedudv}
    \frac{\partial^2 E_{N}(\Delta v)}{\partial U \partial \Delta v}\Big|_{\Delta v = \Delta v^{\bxi}(U=0)} = \frac{\nu}{2t(1+\nu^2)^2}
\end{equation}

To obtain the $U$-derivative of the ensemble potential, note the Lieb maximization principle Eq.~\eqref{eqapp:ens_Liebmax} implies
\begin{equation}
    \frac{\partial F^{\bxi}(n,\Delta v)}{\partial \Delta v}\Big|_{\Delta v = \Delta v^{\bxi}} = 0
\end{equation}

Inserting Eq.~\eqref{eqapp:ens_f_taylor} yields
\begin{equation}
    (\xi_{+}+\xi_{-}) \frac{\partial \epsilon_H(\Delta v)}{\partial \Delta v}\Big|_{\Delta v = \Delta v^{\bxi}} + \left(1-\frac{3 \xi_+ + \xi_-}{2}\right) \frac{\partial E_{N}(\Delta v)}{\partial \Delta v}\Big|_{\Delta v = \Delta v^{\bxi}} + n-1 = 0
\end{equation}
and differentiating again with respect to $U$:
\begin{equation}
    \begin{split}
            & (\xi_{+}+\xi_{-}) \frac{\partial^2 \epsilon_H(\Delta v)}{\partial \Delta v \partial U}\Big|_{\Delta v = \Delta v^{\bxi}} + \left(1-\frac{3 \xi_+ + \xi_-}{2}\right) \frac{\partial^2 E_{N}(\Delta v)}{\partial \Delta v \partial U}\Big|_{\Delta v = \Delta v^{\bxi}} 
            \\
            & + \left[ (\xi_{+}+\xi_{-}) \frac{\partial^2 \epsilon_H(\Delta v)}{\partial \Delta v^2}\Big|_{\Delta v = \Delta v^{\bxi}} + \left(1-\frac{3 \xi_+ + \xi_-}{2}\right) \frac{\partial^2 E_{N}(\Delta v)}{\partial \Delta v^2}\Big|_{\Delta v = \Delta v^{\bxi}} \right] \frac{\partial \Delta v^{\bxi}(n)}{\partial U} = 0
    \end{split}
\end{equation}

Using Eqs.~(\ref{eqapp:KS_orbital_ene},\ref{eqapp:d2edu2},\ref{eqapp:dedudv}) and differentiating polynomial Eq.~\eqref{eqapp:polynomial} with respect to $\Delta v$ twice to compute $\frac{\partial^2 E_{N}(\Delta v)}{\partial \Delta v^2}\Big|_{\Delta v = \Delta v^{\bxi}}$ we finally obtain
\begin{equation} \label{eqapp:dvdu}
    \frac{\partial \Delta v^{\bxi}(n)}{\partial U}\Big|_{U=0} = \frac{\nu}{\sqrt{1+\nu^2}} \frac{2-(\xi_{-}+3\xi_{+})}{2(1-\xi_+)}
\end{equation}

We can now insert Eqs.~(\ref{eqapp:d2edu2},\ref{eqapp:dedudv},\ref{eqapp:dvdu}) in Eq.~\eqref{eqapp:d2fdu2} to get the second order term in the Taylor expansion of the universal functional
\begin{equation}
    \frac{\partial^2 F^{\bxi}(n)}{\partial U^2}\Big|_{U=0} = \frac{2-\xi_--3\xi_+}{16t} \left( \frac{(n-1)^2}{(1-\xi_{+})^2} \frac{1-2\xi_{-}-3\xi_{+}}{1-\xi_{+}} -1 \right) \left[ \frac{(1-\xi_{+})^2-(n-1)^2}{(1-\xi_{+})^2} \right]^{3/2}
\end{equation}
which completes the PT2 approximation to the ensemble Hxc energy.
\begin{equation} \label{eqapp:pt2ehxc}
    E^{\bxi}_{\rm Hxc}(n) \underset{PT2}{\approx} U \frac{\partial F^{\bxi}(n)}{\partial U}\Big|_{U=0} + \frac{U^2}{2} \frac{\partial^2 F^{\bxi}(n)}{\partial U^2}\Big|_{U=0}
\end{equation}

\subsection{High density limit and the PADE interpolant}\label{sec:PADE_Hubbard}

To derive the high correlation limit corresponding to $u=U/2t \rightarrow \infty$, we reexpress the polynomial Eq.~\eqref{eqapp:polynomial} with $\tilde{E}_{N,i} = E_{N,i}/U$, $\Delta \tilde{v} = \Delta v / U$, and $u$:
\begin{equation} \label{eqapp:polynomial_u}
    \tilde{E}_{N,i}^3 = -\frac{1}{u^2} + \left(\frac{1}{u^2} - 1 + \Delta \tilde{v}^2 \right) \tilde{E}_{N,i} + 2 \tilde{E}_{N,i}^2
\end{equation}
in the high-$u$ limit, it becomes
\begin{equation} \label{eqapp:polynomial_uinfty}
    \tilde{E}_{N,i}^3 \underset{u\rightarrow\infty}{=} \left(\Delta \tilde{v}^2 - 1 \right) \tilde{E}_{N,i} + 2 \tilde{E}_{N,i}^2
\end{equation}
which can straightforwardly be put under the form of a second order polynomial, with lowest solution (i.e. the ground-state) given by
\begin{equation}
    \tilde{E}_{N} \underset{u\rightarrow\infty}{=} {\rm min} \big\{ 0, 1 - \Delta \tilde{v} \big\}
\end{equation}
This, together with the knowledge of $u \rightarrow \infty$ limit for the non-interacting one-electron energy,
\begin{equation}
    \frac{\epsilon_H(\Delta \tilde{v})}{U} \underset{u\rightarrow\infty}{=} - \left| \frac{\Delta \tilde{v}}{2} \right|
\end{equation}
allows to rexpress the ensemble functional of Eq.~\eqref{eqapp:ens_f_taylor}
\begin{equation}
    \frac{F^{\bxi}(n)}{U} \underset{u\rightarrow\infty}{=} \underset{\Delta \tilde{v}}{{\rm sup}} \left\{ \xi_+ - |\Delta \tilde{v}| \frac{\xi_+ + \xi_-}{2} + \left(1-\frac{3\xi_+ + \xi_-}{2}\right) {\rm min}\big\{ 0, 1 - \Delta \tilde{v} \big\} + \Delta v (n-1) \right\}
\end{equation}

To make further progress, one has to consider several $\Delta \tilde{v}$ ranges separately. For $|\Delta \tilde{v}|\leq1$,
\begin{equation}
    \frac{F^{\bxi}(n)}{U} \underset{u\rightarrow\infty, |\Delta \tilde{v}|\leq1}{=} \underset{|\Delta \tilde{v}|\leq1}{{\rm sup}} \left\{ \xi_+ - |\Delta \tilde{v}| \frac{\xi_+ + \xi_-}{2}  + \Delta v (n-1) \right\}
\end{equation}
which is further decomposed into
\begin{equation}
    \frac{F^{\bxi}(n)}{U} \underset{u\rightarrow\infty, |\Delta \tilde{v}|\leq1}{=} \left\{\begin{split}
         \quad & {\rm sup} \left\{ \xi_+ \,,\, \frac{\xi_+ - \xi_-}{2}  -(n-1) \right\} \quad {\rm if} \; -1 \leq|\Delta \tilde{v}|\leq 0
        \\
        & {\rm sup} \left\{ \xi_+ \,,\,\frac{\xi_+ - \xi_-}{2}  +(n-1) \right\}  \quad {\rm if} \; 0 \geq|\Delta \tilde{v}|\geq 1
    \end{split} \right.    
\end{equation}
so that we can finally state
\begin{equation}
    \frac{F^{\bxi}(n)}{U} \underset{u\rightarrow\infty, |\Delta \tilde{v}|\leq1}{=} {\rm sup} \left\{ \xi_+ \,,\, \frac{\xi_+ - \xi_-}{2} + |n-1| \right\}
\end{equation}
Shifting to $|\Delta \tilde{v}|\geq1$,
\begin{equation}
    \frac{F^{\bxi}(n)}{U} \underset{u\rightarrow\infty, |\Delta \tilde{v}|\geq 1}{=} \underset{|\Delta \tilde{v}|\geq 1}{{\rm sup}} \left\{ \xi_+ - |\Delta \tilde{v}| \frac{\xi_+ + \xi_-}{2} + \left(1-\frac{3\xi_+ + \xi_-}{2}\right) \big(1 - \Delta \tilde{v} \big) + \Delta v (n-1) \right\}
\end{equation}
we first consider negative values
\begin{equation}
    \frac{F^{\bxi}(n)}{U} \underset{u\rightarrow\infty}{=} \underset{\Delta \tilde{v}\leq -1}{{\rm sup}} \Big\{\Delta \tilde{v} \big(n-\xi_+\big)  \Big\} + 1 - \frac{\xi_+ + \xi_-}{2} \quad {\rm if} \; \Delta \tilde{v}\leq -1
\end{equation}
To make progress, we take insight from non-interacting $v$-representability: the analytical expression of $T^{\bxi}_s(n)$ reads
\begin{equation} \label{eqapp:ens_Ts_Liebmax}
    \begin{split}
            T^{\bxi}_s(n) & = \underset{\Delta v}{\rm sup} \Big\{ \xi_- \, \mathcal{E}_{N_-} (\Delta v) + \xi_+ \, \mathcal{E}_{N_+} (\Delta v) + \left(1-\frac{3\xi_+ + \xi_-}{2}\right) \, \mathcal{E}_{N} (\Delta v) + \Delta v (n-1) \Big\}
            \\
            & = \underset{\Delta v}{\rm sup} \Big\{ (\xi_- + \xi_+) \, \epsilon_{H} (\Delta v) + \xi_+ \, U + \left(1-\frac{3\xi_+ + \xi_-}{2}\right) \, \epsilon_{H} (\Delta v) + \Delta v (n-1) \Big\}
            \\
            & = \underset{\Delta v}{\rm sup} \Big\{ -2  \sqrt{t^2 + (\Delta v/2)^2} \, (1-\xi_+) + \Delta v (n-1) \Big\}
    \end{split}
\end{equation}
inserting the maximizing potential expression, that is KS potential Eq.~\eqref{eqapp:Dvs_n}, one finally gets
\begin{equation}
    T^{\bxi}_s(n) = -2 t \sqrt{(1-\xi_+)^2-(n-1)^2}
\end{equation}
which is well-defined only when
\begin{equation} \label{eqapp:nonvrep}
    1-\xi_+ \geq n-1 \geq \xi_+ -1
\end{equation}
Thus, $n-\xi_+ \geq 0$ and
\begin{equation}
    \frac{F^{\bxi}(n)}{U} \underset{u\rightarrow\infty}{=} \frac{\xi_+ - \xi_-}{2} -(n-1) \quad {\rm if} \; \Delta \tilde{v}\leq -1
\end{equation}
Moving to $\Delta \tilde{v}\geq 1$, we have
\begin{equation}
    \frac{F^{\bxi}(n)}{U} \underset{u\rightarrow\infty}{=} \underset{\Delta \tilde{v}\leq -1}{{\rm sup}} \Big\{\Delta \tilde{v} \big(n-2+\xi_+\big)  \Big\} + 1 - \frac{\xi_+ + \xi_-}{2} \quad {\rm if} \; \Delta \tilde{v}\geq 1
\end{equation}

As from Eq.~\eqref{eqapp:nonvrep}, we have $n-2+\xi_+ \leq 0$, we obtain
\begin{equation}
    \frac{F^{\bxi}(n)}{U} \underset{u\rightarrow\infty}{=} \frac{\xi_+ - \xi_-}{2} +(n-1) \quad {\rm if} \; \Delta \tilde{v}\geq 1
\end{equation}
which can be rewritten
\begin{equation}
    F^{\bxi}(n) \underset{u\rightarrow\infty}{=} U \; {\rm sup} \left\{ \xi_+ \,,\, \frac{\xi_+ - \xi_-}{2} + |n-1| \right\}
\end{equation}

Knowing the high-$U$ limit and the low-$U$ second-order Taylor expansion, we can build the following Padé approximant to interpolate between them:
\begin{equation} \label{eqapp:pade1}
    E_{\rm Hxc}^{\bxi}(n) \approx a^{\bxi}(n) U +  \frac{b^{\bxi}(n) U^2}{1+c^{\bxi}(n) U}
\end{equation}
where all terms are given by
\begin{equation} \label{eqapp:pade2}
    \left\{ \begin{split}
        \; a^{\bxi}(n) & = \frac{\partial F^{\bxi}(n)}{\partial U}\Big|_{U=0}
        \\
        b^{\bxi}(n) & = \frac{1}{2} \frac{\partial^2 F^{\bxi}(n)}{\partial U^2}\Big|_{U=0}
        \\
        c^{\bxi}(n) & = \frac{b^{\bxi}(n)}{\gamma^{\bxi}(n)-a^{\bxi}(n)}
        \\
        \gamma^{\bxi}(n) & = {\rm sup} \left\{ \xi_+ \,,\, \frac{\xi_+ - \xi_-}{2} + |n-1| \right\}
    \end{split} \right.
\end{equation}

\section{Numerical details and complementary results}

\subsection{Computational details} \label{sec:comp_details}

All calculations are done at the exact ensemble density $n^{\bxi}$ determined from the system parameters using Eqs.~\eqref{eqapp:ind_dens} to~\eqref{eqapp:ens_density_pot}. See definitions in and below Eq.~\eqref{eqapp:costheta}. Of course, calculations using scaling functions are performed in the zero-weight limit where the ensemble density reduces to the 2-electron ground state density given in Eq.~\eqref{eqapp:ind_dens}. Various values of $U$ and $\Delta v$ will be considered, but in all cases we set $t=1$.

\bigskip
{\bf Calculations using full-ensemble functionals} -- For arbitrary weights $(\xi_-,\xi_+)$, Fukui functions are computed from the working equation:
\begin{equation} \label{eqapp:fukui}
    \begin{split}
        f_{\pm} 
        &=  \big(1+\chi^{\bxi} f_{\rm Hxc}^{\bxi}\big) f^{\bxi}_{s,\pm} - \chi^{\bxi} f_{\rm Hxc}^{\bxi} \frac{n^{\bxi}}{2} 
        \\
        &\quad  \pm \; \chi^{\bxi} \sum_{\lambda=+,-} \Big( \delta_{\lambda\pm} \mp\frac{\xi_{\lambda}}{2}   \Big) \frac{\partial v_{\rm Hxc}^{\bxi}(n)}{\partial \xi_{\lambda}}\Big|_{n=n^{\bxi}}. 
    \end{split}
\end{equation} 
with the shorthand notation $f^{\bxi}:=f^{\bxi}(n^{\bxi})$ for any function of the density (one site 0). The KS Fukui functions $f^{\bxi}_{s,\pm}$ are computed analytically from the ensemble density using Eq.~\eqref{eqapp:ks_fukui}. The ensemble response $\chi^{\bxi}$ is computed from ensemble Dyson Eq.~\eqref{eqapp:ens_dyson} using the analytical KS response defined through Eq.~\eqref{eqapp:ens_ks_resp} and the Hxc kernel $f^{\bxi}_{\rm Hxc}$. The latter is evaluated through analytical differentiation of the relevant density functional approximation of Hxc energy. Weight-derivatives of the Hxc potential are obtained in the same manner.

Note however that our SOFT convention to express everything as a function of density on site 0 brings some sign changes compared to general EDFT expressions: from the first line of Eq.~\eqref{eqapp:ens_f_taylor} and its non-interacting analog, it is clear our external and KS potentials have a flipped sign compared to general EDFT expressions using functionals
\begin{equation} \label{eqapp:pots_from_funcs}
    \left\{\begin{split}
        \Delta v^{\bxi}_s(n) & = \frac{\partial T^{\bxi}_s(n)}{\partial n}
        \\
        \Delta v^{\bxi}(n) & = \frac{\partial F^{\bxi}(n)}{\partial n}
    \end{split}\right.
\end{equation}
From the Hxc energy and potential definitions through Eq.~\eqref{eqapp:defEhxc}, which carries straightforwardly to SOFT, and through Eq.\eqref{eqapp:defSOFTvhxc} respectively, we see the sign flip carries to the relation between ensemble Hxc energy and potential:
\begin{equation} \label{eqapp:vhxcfromehxc}
    \begin{split}
        \Delta v^{\bxi}_{\rm Hxc}(n) & = \Delta v^{\bxi}_s(n) - \Delta v^{\bxi}(n)
        \\
        & = \frac{\partial}{\partial n} \big[T^{\bxi}_s(n)-F^{\bxi}(n)\big]
        \\
        & = - \frac{\partial E^{\bxi}_{\rm Hxc}(n)}{\partial n}
    \end{split}
\end{equation}

Regarding the kernel, owing to the KS and physical response function definitions Eqs.~(\ref{eqapp:ens_ks_resp},\ref{eqapp:ens_resp_softdef}), differentiating the above expression for the Hxc potential with respect to $n$ and using ensemble Dyson Eq.~\eqref{eqapp:ens_dyson} gives
\begin{equation}
    f^{\bxi}_{\rm Hxc}(n) = \frac{\partial\Delta v^{\bxi}_{\rm Hxc}(n)}{\partial n} = - \frac{\partial^2 E^{\bxi}_{\rm Hxc}(n)}{\partial n^2}
\end{equation}

\bigskip
{\bf Calculations using scaling functions} -- The exact ground-state functional is always used in combination with scaling functions, everything being evaluated at the exact ground-state density. 
All scaling functions are evaluated from relevant Hxc potential approximations at the given weights $\bxi$ and for 0-weights.
\rev{\begin{equation}
    s_{\rm Hx}(\bxi) = \frac{v^{\bxi}_{\rm Hx}[n^{\bxi}]}{v_{\rm Hx}[n^{\bxi}]}\,, \qquad s_{\rm c}[n](\bxi) = \frac{v^{\bxi}_{\rm c}[n]}{v_{\rm c}[n]}\,, \qquad v^{\bxi}_{\rm Hxc}[n] = s_{\rm Hx}(\bxi) v_{\rm Hx}[n] + s_{\rm c}[n](\bxi) v_{\rm c}[n]
\end{equation}}
We stress again \rev{that Hx} potentials are to be evaluated at the relevant ensemble density $n^{\bxi}$ to build the scaling function.
The purely Hx scaling (or EEXX scaling) is computed from the ratio of Hx potentials (obtained by inserting Eq.~\eqref{eqapp:ens_ehx} in Eq.~\eqref{eqapp:vhxcfromehxc}). At the EEXX level for the scaling function, the density dependences simplify out. 
For the double-scaling at PT2 level, the Hx scaling function is given by EEXX potential, while the correlation scaling $s_c(\bxi)$ is built from the ratio of correlation potentials obtained by differentiating the second term of Eq.~\eqref{eqapp:pt2ehxc} with respect to $n$ (see again Eq.~\eqref{eqapp:vhxcfromehxc}).

\subsection{EEXX and PT2 in the low correlation regime}

\begin{figure}[h]
    \centering
    \includegraphics[width=0.5\linewidth]{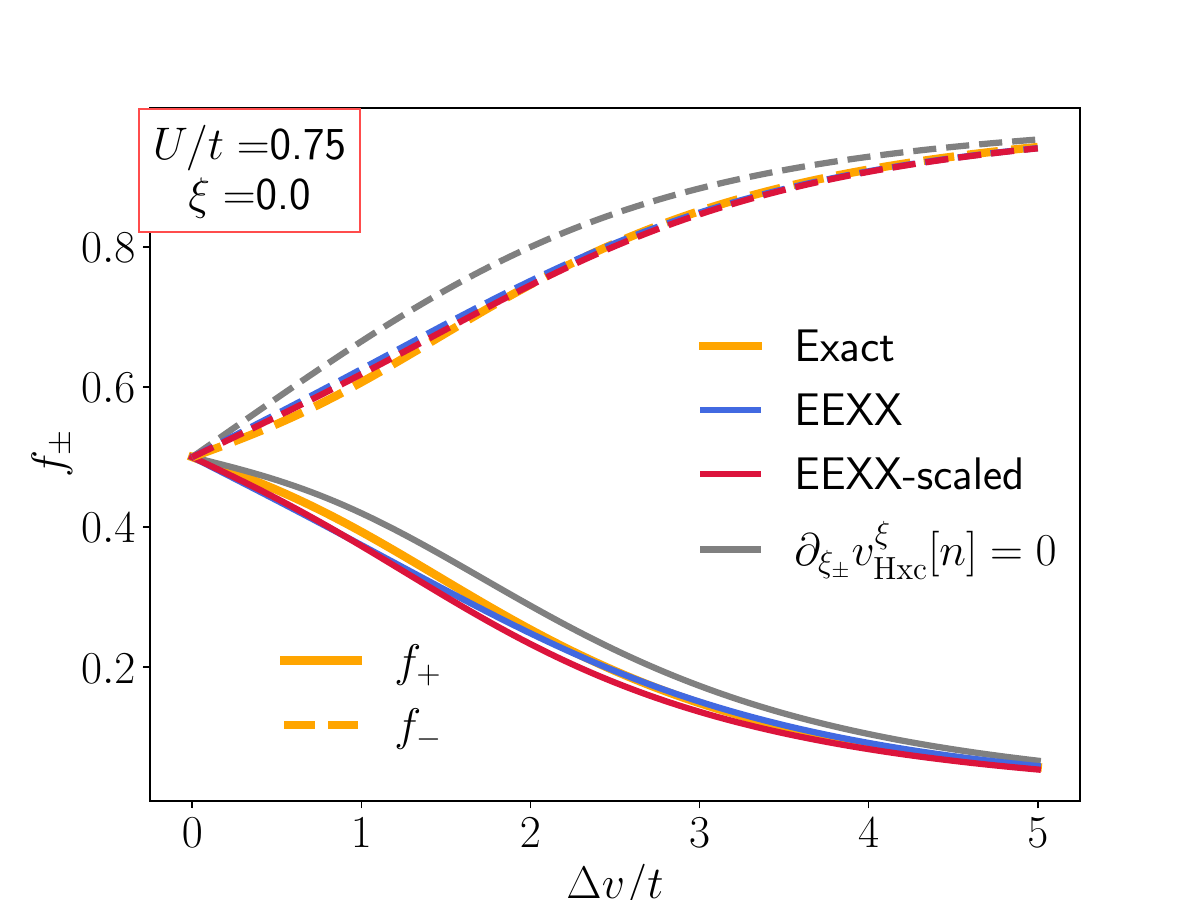}
    \caption{Fukui functions in weak correlation regime for null ensemble weights (parameters in the red inset). Full and dashed lines are $f_{+}$ and $f_{-}$ calculations respectively. Exact results (yellow lines) are compared to EEXX (blue lines) and to the neglect of weight-derivative contributions (grey lines).}
    \label{fig:EEXXlowU}
\end{figure}

When correlation is low, one expects EEXX to capture most of the physics. Its performance in predicting Fukui functions in this regime is shown in Fig.~\ref{fig:EEXXlowU} for $U=0.75$ and 0 weights. It is in good agreement with exact results and the neglect of weight-derivative terms in Eq.~\eqref{eqapp:fukui} is shown for comparison. The latter result uses the exact ground-state functional to evaluate exactly all other terms.

\begin{figure}[h]
    \centering
    \includegraphics[width=0.5\linewidth]{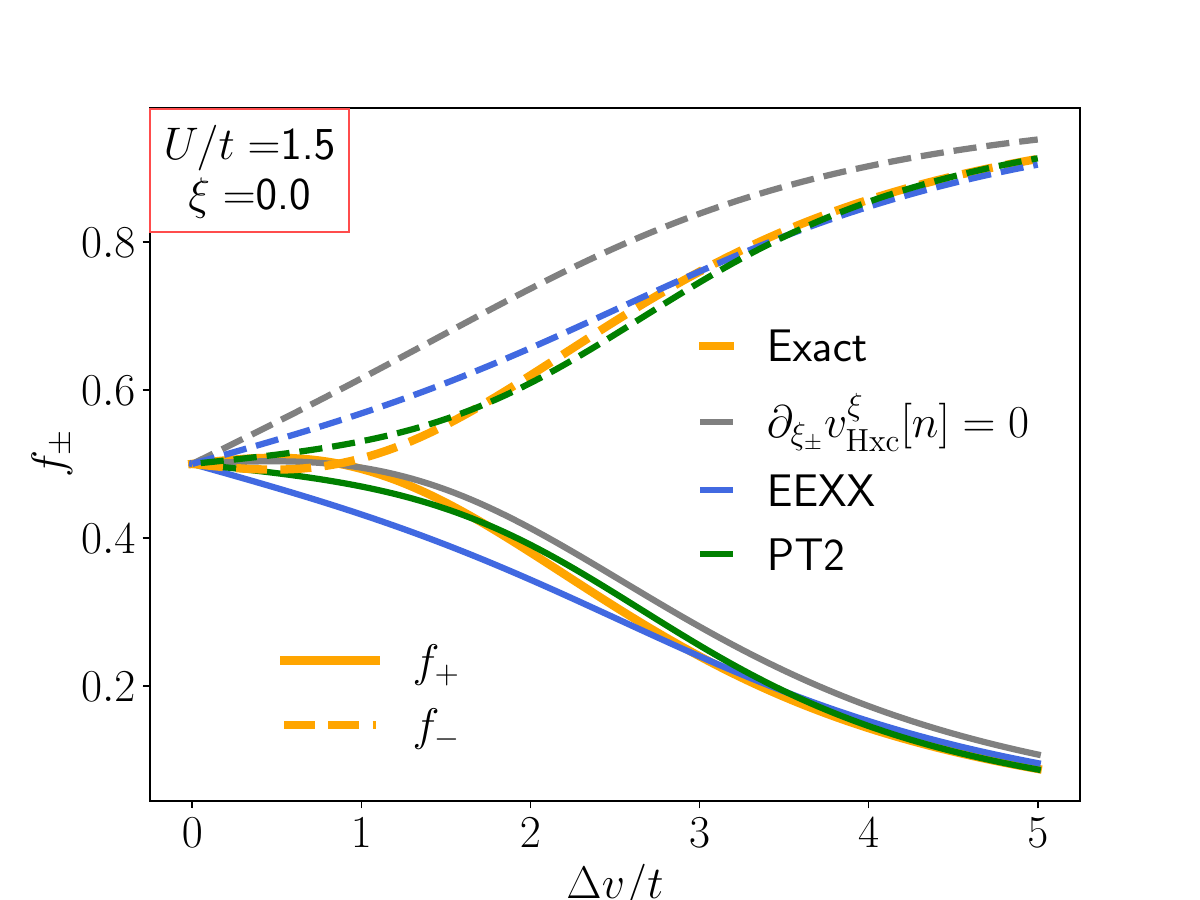}
    \caption{Fukui functions in the low correlation regime for null ensemble weights (parameters in the red inset). We compare EEXX ensemble DFA (blue) to PT2 (green) and neglect of weight derivative terms (grey). Exact results are shown in orange.}
    \label{fig:PT2midU}
\end{figure}

The situation changes when increasing correlation up to $U=1.5$ (keeping 0 weights) as shown on Fig.~\ref{fig:PT2midU}. A noticeable discrepancy between EEXX and exact results appears for low external potentials $\Delta v < 2U$, where the correlation energy is the highest. Note however the improvement over neglect of weight-derivative terms is still noticeable for $f_-$. Inclusion of correlation in the PT2 DFA allows to recover a close agreement with exact results in all the $\Delta v$ range.

\subsection{Scaling functions}
{\bf Inadequacy of an Hxc energy-based scaling function} -- Using the ratio of Hxc ensemble and ground-state energies at the ensemble density \rev{to scale the ground-state potential} is a poor strategy to predict the Fukui function.

\begin{equation}
    s(\bxi) = \frac{E^{\bxi}_{\rm Hxc}[n^{\bxi}]}{E_{\rm Hxc}[n^{\bxi}]}\,, \qquad v_{\rm Hxc}^{\bxi}[n] = s(\bxi) E_{\rm Hxc}[n]
\end{equation}

\begin{figure}[H]
    \centering
    \includegraphics[width=0.5\linewidth]{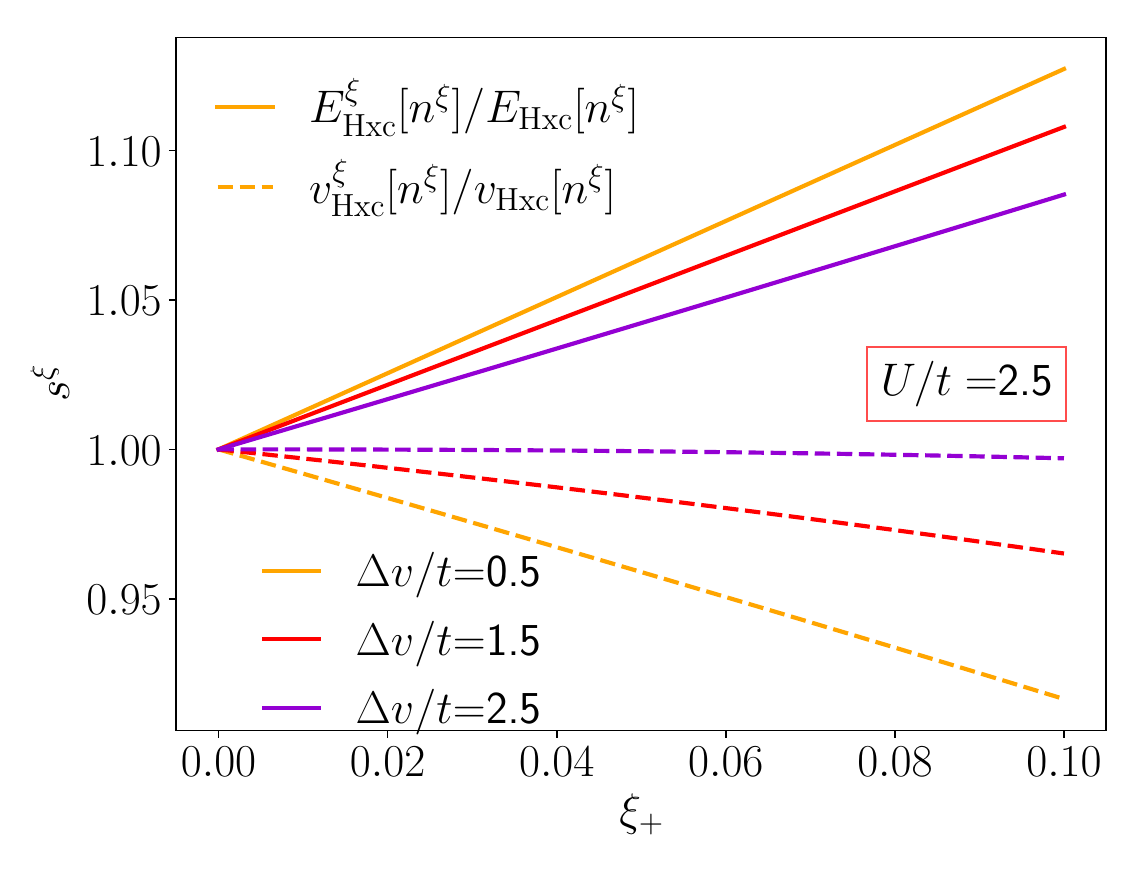}
    \caption{Profiles of the scaling function as a function of $\xi_+$ defined from the Hxc energy ratio (full lines) compared to the ratio of potentials (dashed lines). The weight $\xi_-$ is kept at 0. Three values of the external potential are considered.}
    \label{fig:s_v_vs_e}
\end{figure}

The reason is that they encode widely-different weight-dependence, as seen on Fig.~\ref{fig:s_v_vs_e} where we compare the two definitions for the scaling function at $U/t=2.5$. The weight $\xi_-$ is kept fixed at 0, while $\xi_+$ is varied. Different values of the external potential, producing different ensemble densities, are considered. In all cases, it is seen that the scaling functions using energies and potentials vary in an almost opposite way, making the former unsuited to encode weight-dependence of the Hxc potential.

\begin{figure}[H]
    \centering
\begin{tabular}{cc}
    \includegraphics[width=0.47\linewidth]{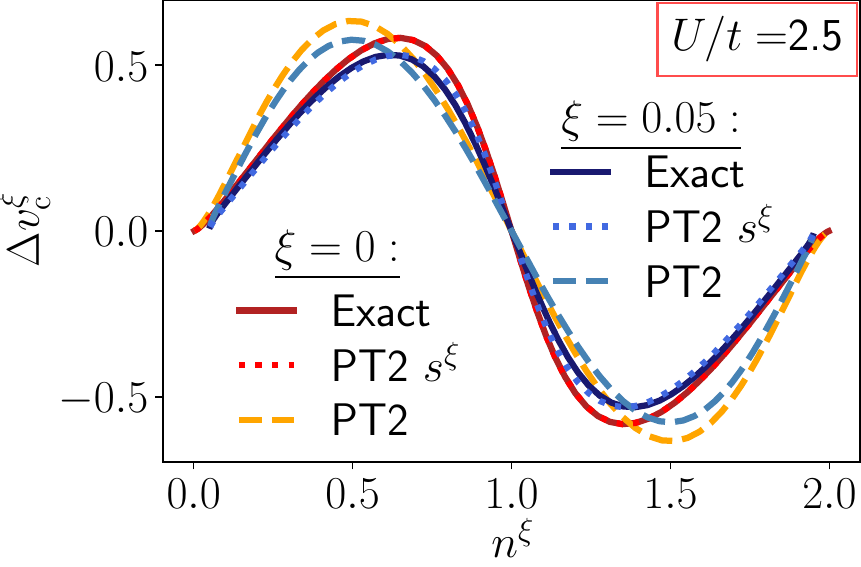}&
    \includegraphics[width=0.4525\linewidth]{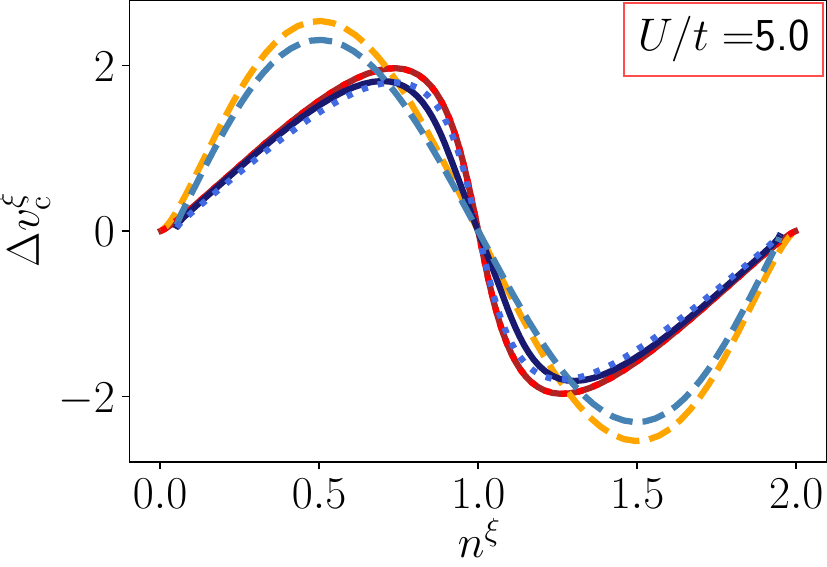}\\[2\tabcolsep]
    \includegraphics[width=0.47\linewidth]{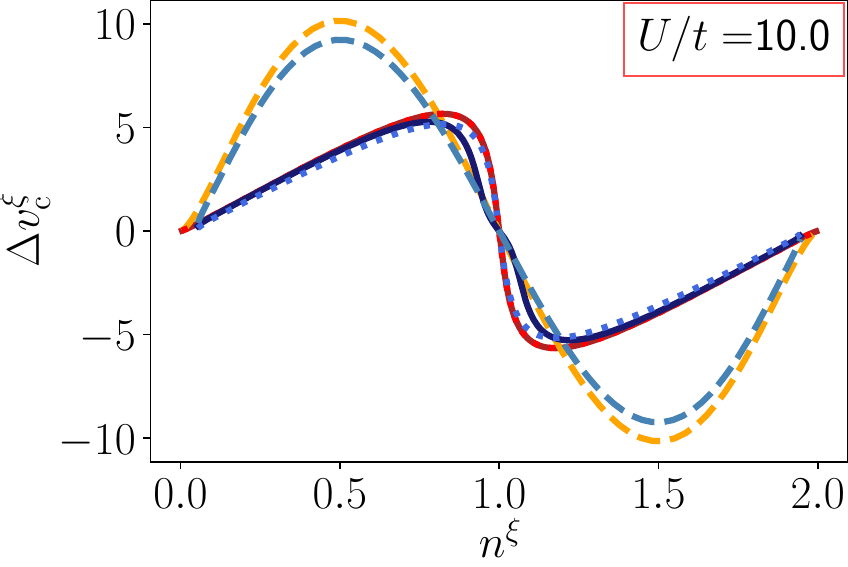}&
    \includegraphics[width=0.455\linewidth]{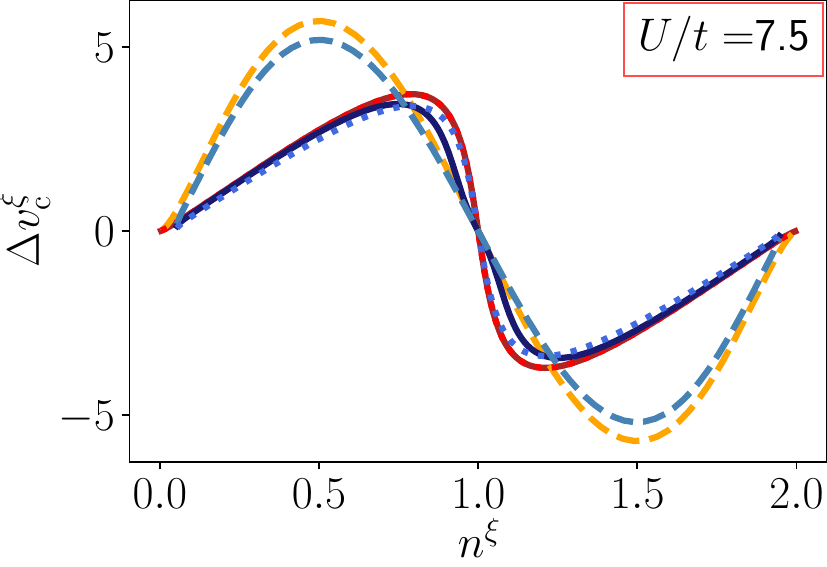}
\end{tabular}
    \caption{Ensemble correlation energy as a function of ensemble density. Exact results (full lines) are compared to full PT2 potential (long-dashed lines) and to the PT2 double scaling function applied to the ground state potential (short-dashed lines). Two values are considered for the weights, taken to be equal $\xi=\xi_+=\xi_-$. Each panel corresponds to a specific value of $U$. }
    \label{fig:vc_pt2scaling}
\end{figure}

\bigskip
{\bf Improvement brought by double scaling at the PT2 level} -- To understand the impact of the double-scaling, we compare on Fig.~\ref{fig:vc_pt2scaling} the ensemble correlation potential predicted by the full-PT2 DFA to the one obtained by scaling the ground-state correlation potential using to the ratio of PT2 ensemble and ground-state correlation potentials. Each of the four panels correspond to a different correlation strength $U/t$. On each panel, in the 0-weight case, the scaling strategy is exact by construction, while the full PT2 correlation potential exhibits increasing errors with $U/t$ when the system is strongly asymmetric. Results for $\xi=\xi_+=\xi_-=0.05$ show the scaled potential matches closely the exact ensemble one, even for large $U/t$, with the brunt of the error around the symmetric case. This shows the weight-derivatives of the Hxc potential will be modeled more accurately for large $\Delta v$ thanks to the scaling approach, while they will not improve nor deteriorate results around $\Delta v = 0$. 

\begin{figure}[H]
    \centering
    \includegraphics[width=0.5\linewidth]{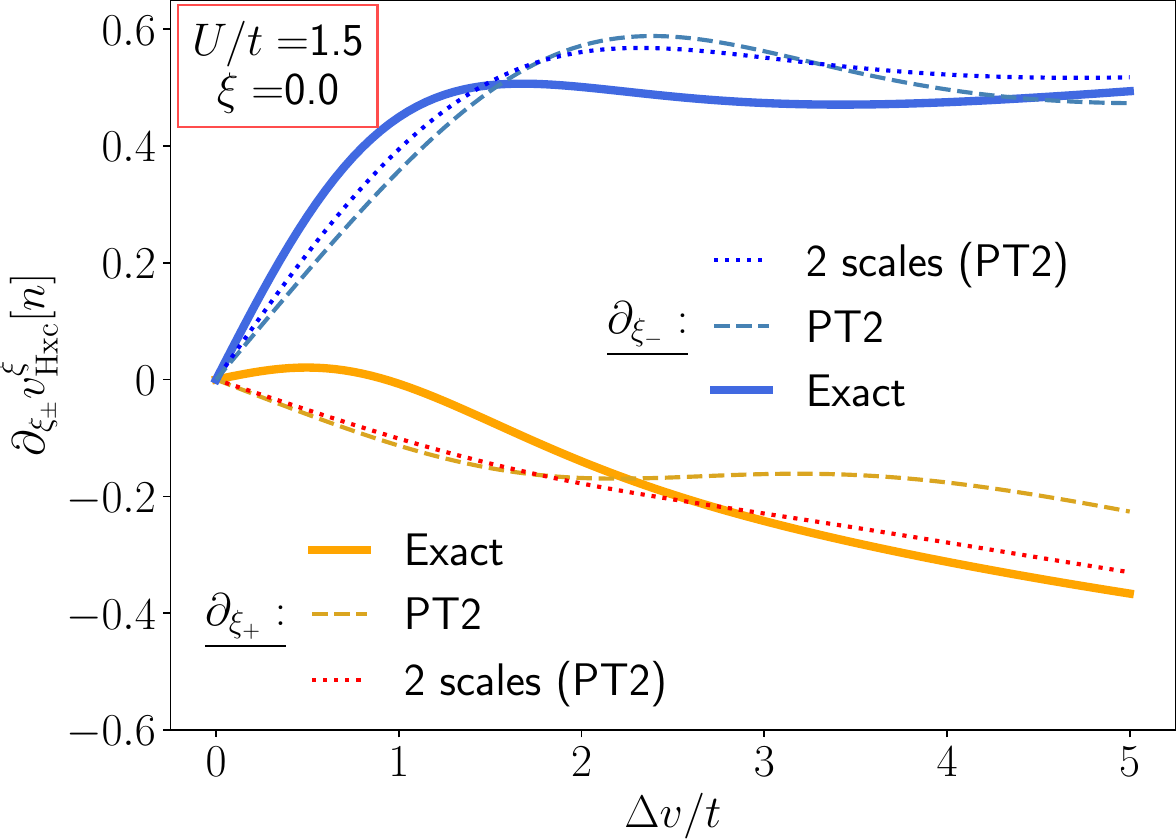}
    \caption{Weight-derivatives of the ensemble Hxc potential in the 0-weight limit. Exact results (full lines) are compared to full PT2 EDFA (long-dashed lines) and PT2 scaling of ground state DFA (short-dashed lines).}
    \label{fig:dxiv_s_PT2vsEEXX1.5}
\end{figure}

This is illustrated on Fig.~\ref{fig:dxiv_s_PT2vsEEXX1.5} for $U/t=1.5$, comparing exact weight-derivatives of the ensemble Hxc energy to the full PT2 EDFA and the PT2-derived double scaling.

\subsection{Smoothing of Pade interpolant}

\begin{figure}[h]
    \centering
    \begin{tabular}{ccc}
    \includegraphics[width=0.32\linewidth]{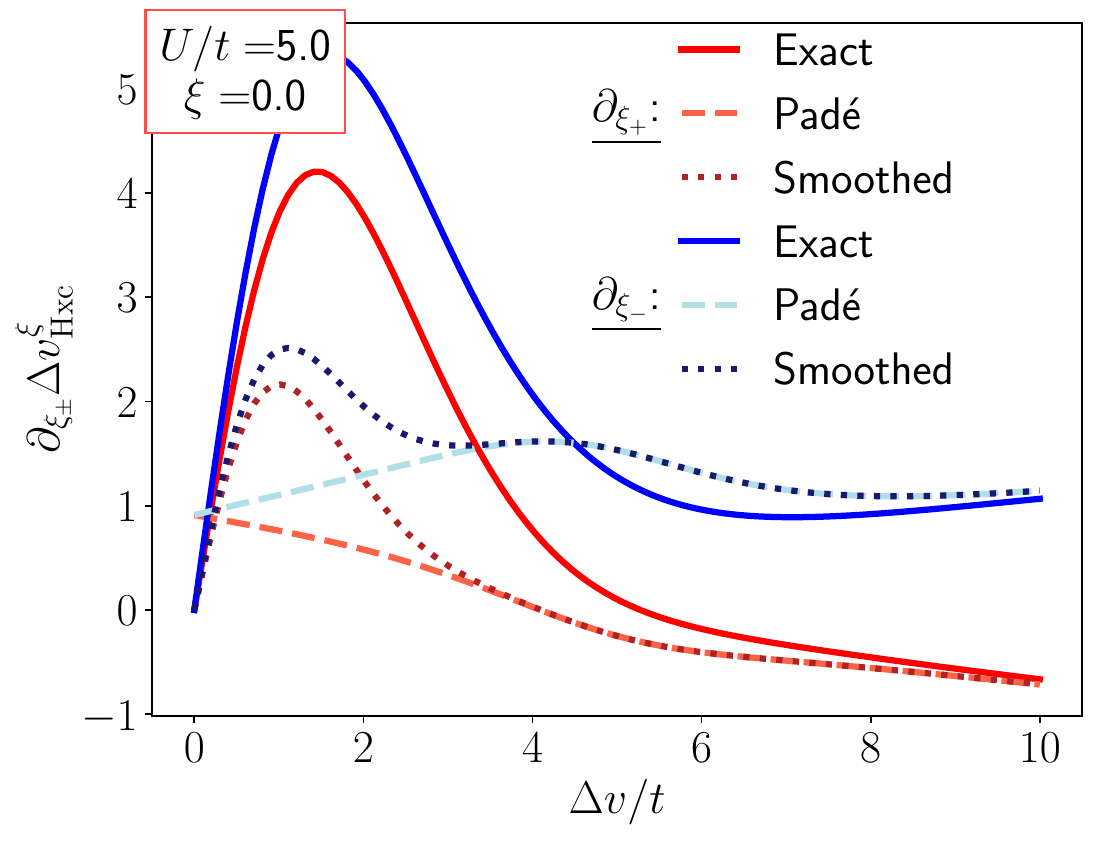}&
    \includegraphics[width=0.32\linewidth]{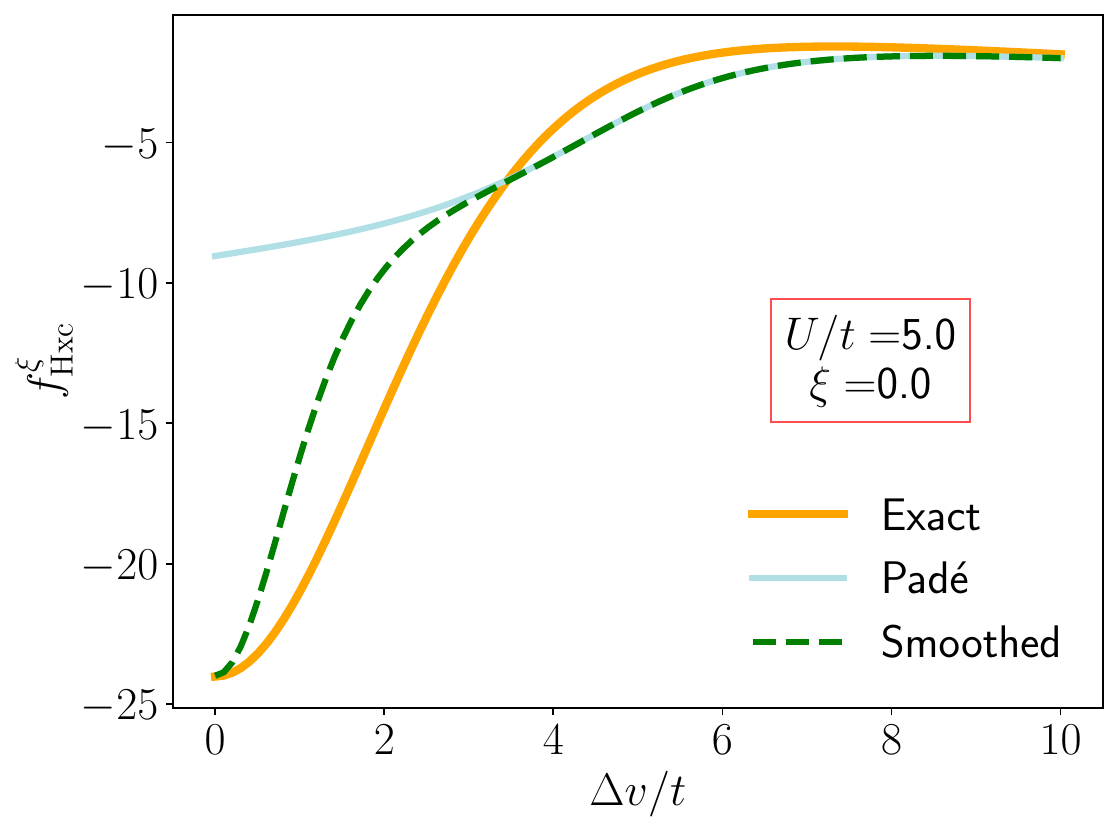}&
    \includegraphics[width=0.32\linewidth]{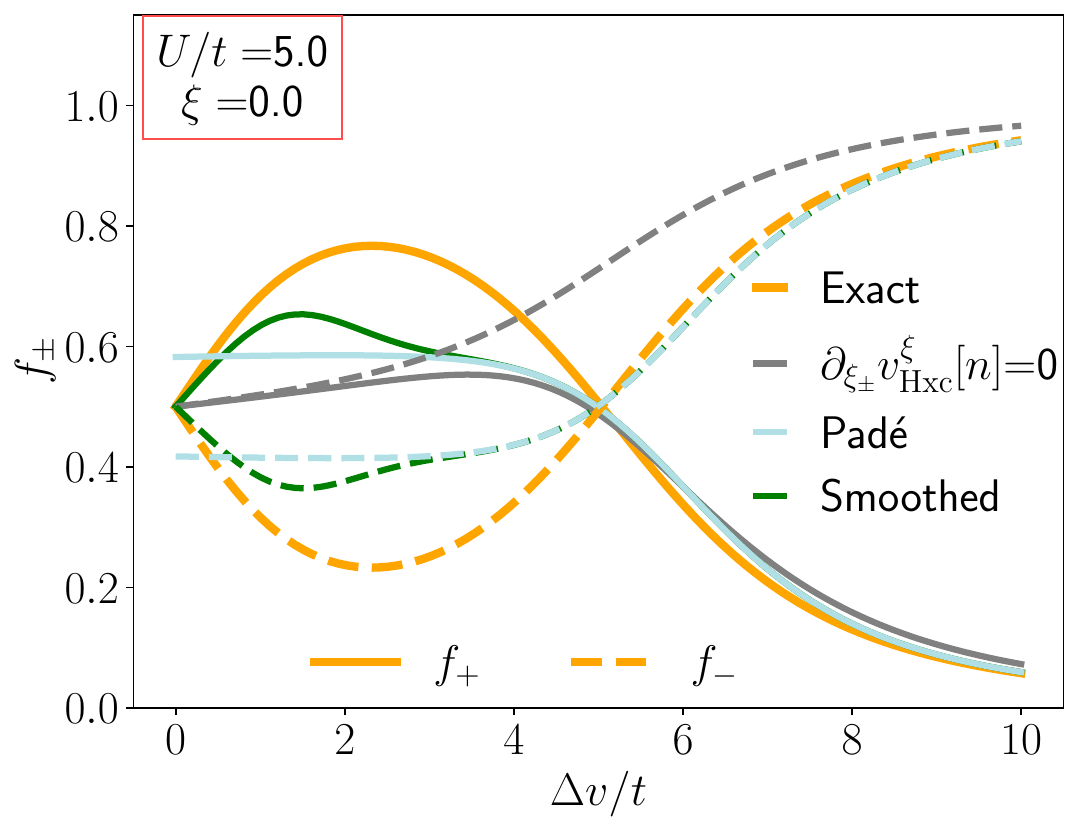}
    \end{tabular}
    \caption{Effect of smoothing the strictly-correlated limit in the Padé approximant. Left panel: Weight-derivatives of the Hxc potential. Middle panel: Hxc kernel. Right panel: Fukui functions.}
    \label{fig:spvspfU5x0}
\end{figure}

\begin{figure}[h]
    \centering
    \begin{tabular}{ccc}
    \includegraphics[width=0.32\linewidth]{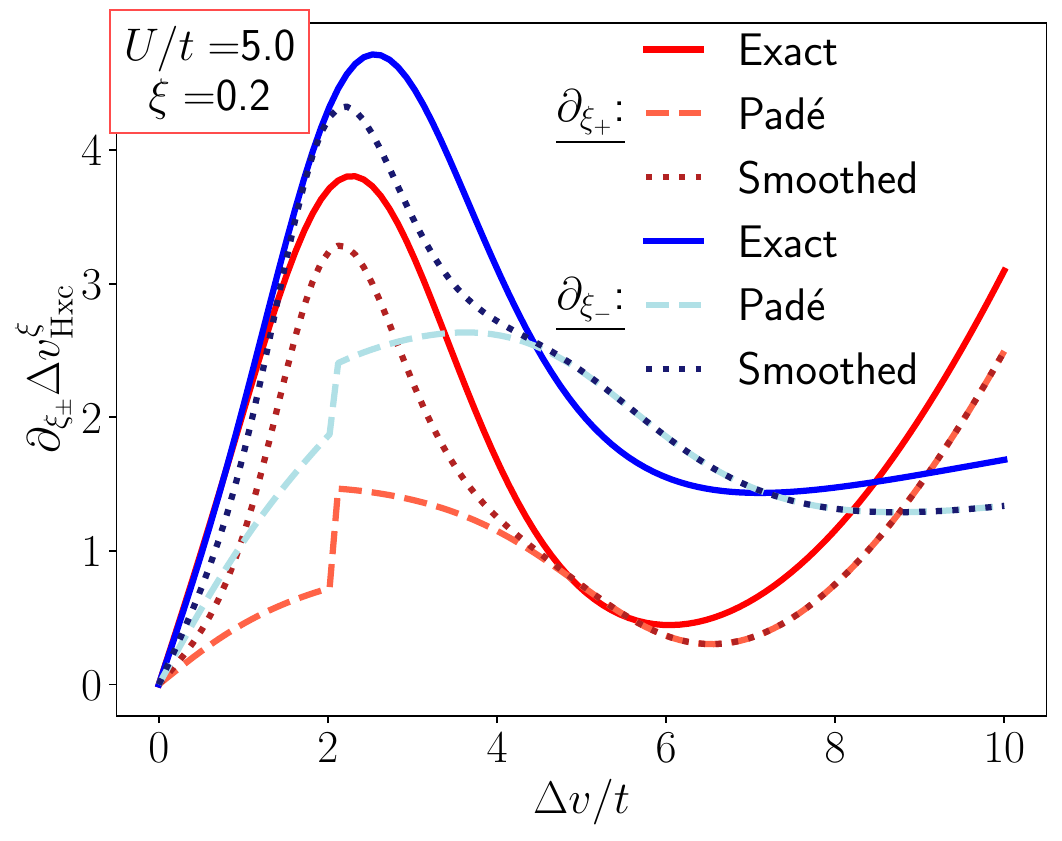}&
    \includegraphics[width=0.32\linewidth]{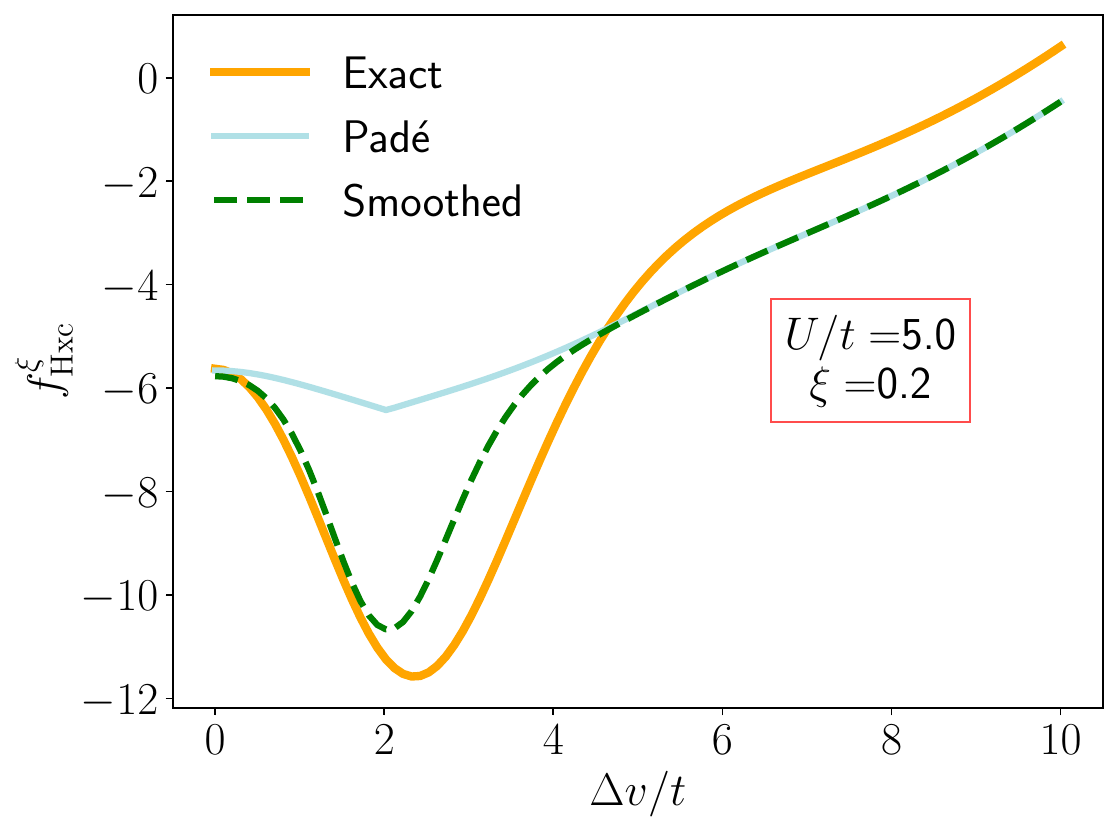}&
    \includegraphics[width=0.32\linewidth]{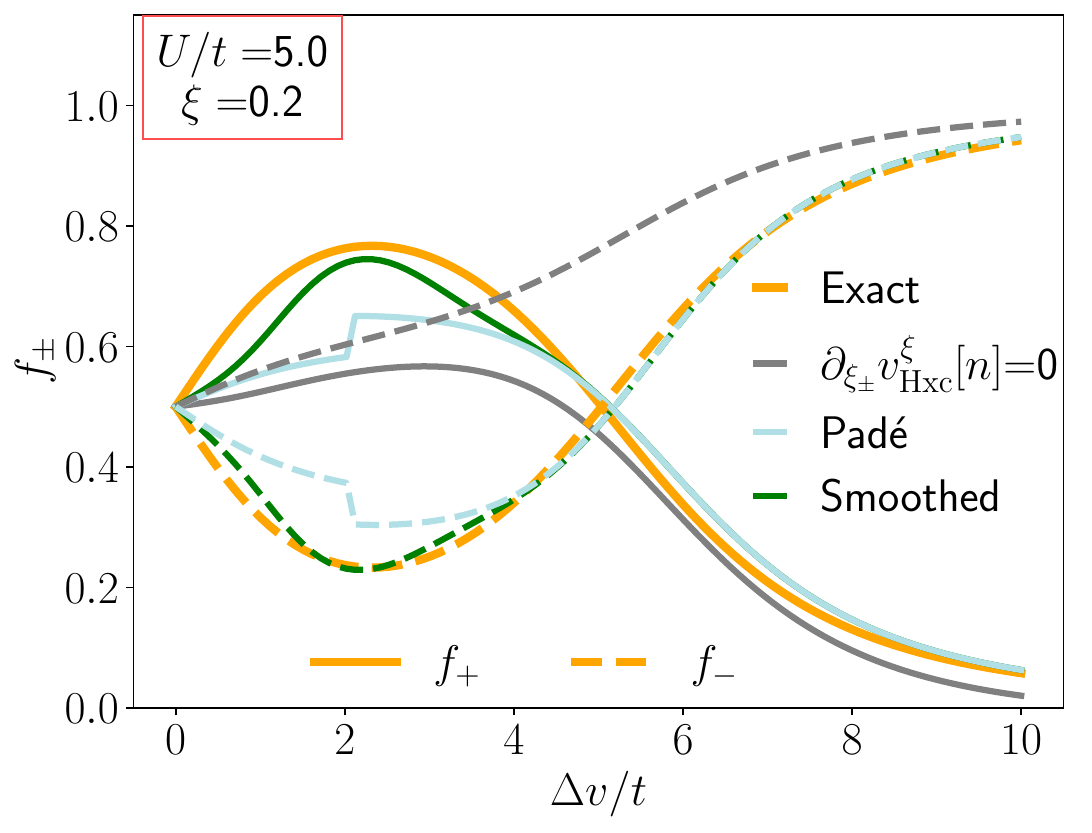}
    \end{tabular}
    \caption{Same as Fig.~\ref{fig:spvspfU5x0} for non-zero weights $\xi=\xi_+=\xi_-$.}
    \label{fig:spvspfU5x2}
\end{figure}

\begin{figure}[h]
    \centering
    \begin{tabular}{ccc}
    \includegraphics[width=0.32\linewidth]{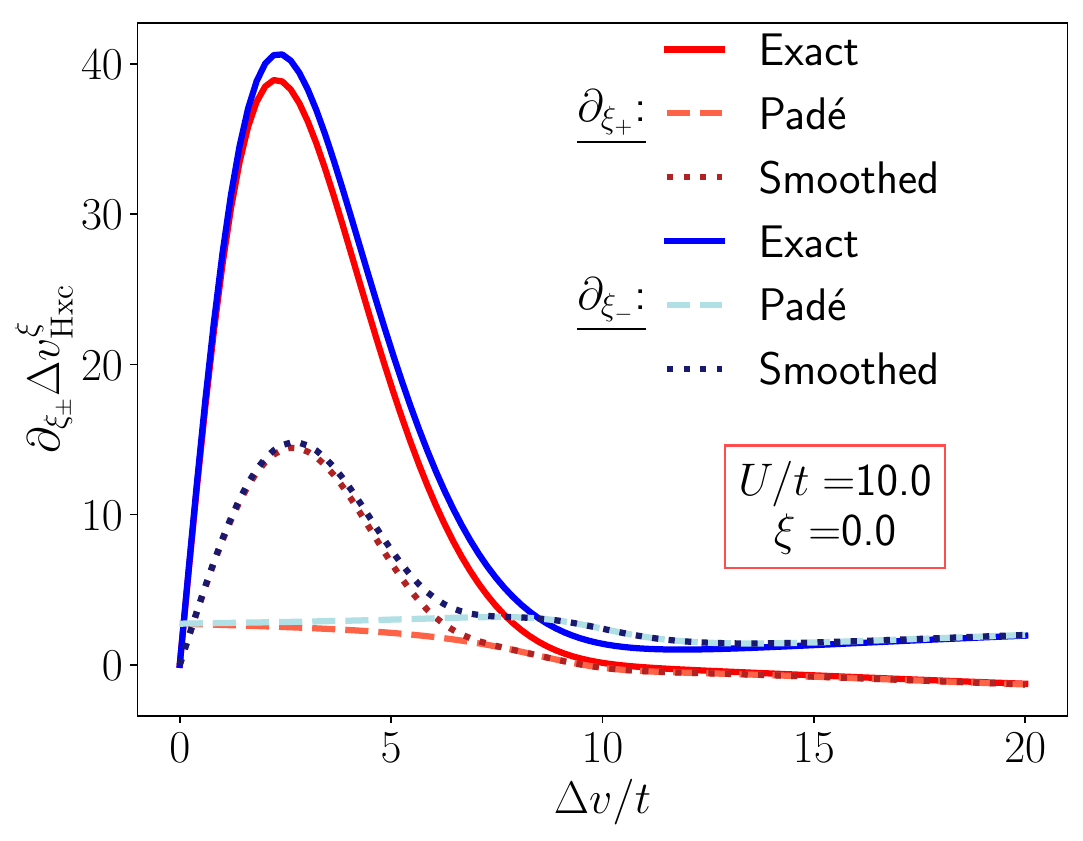}&
    \includegraphics[width=0.32\linewidth]{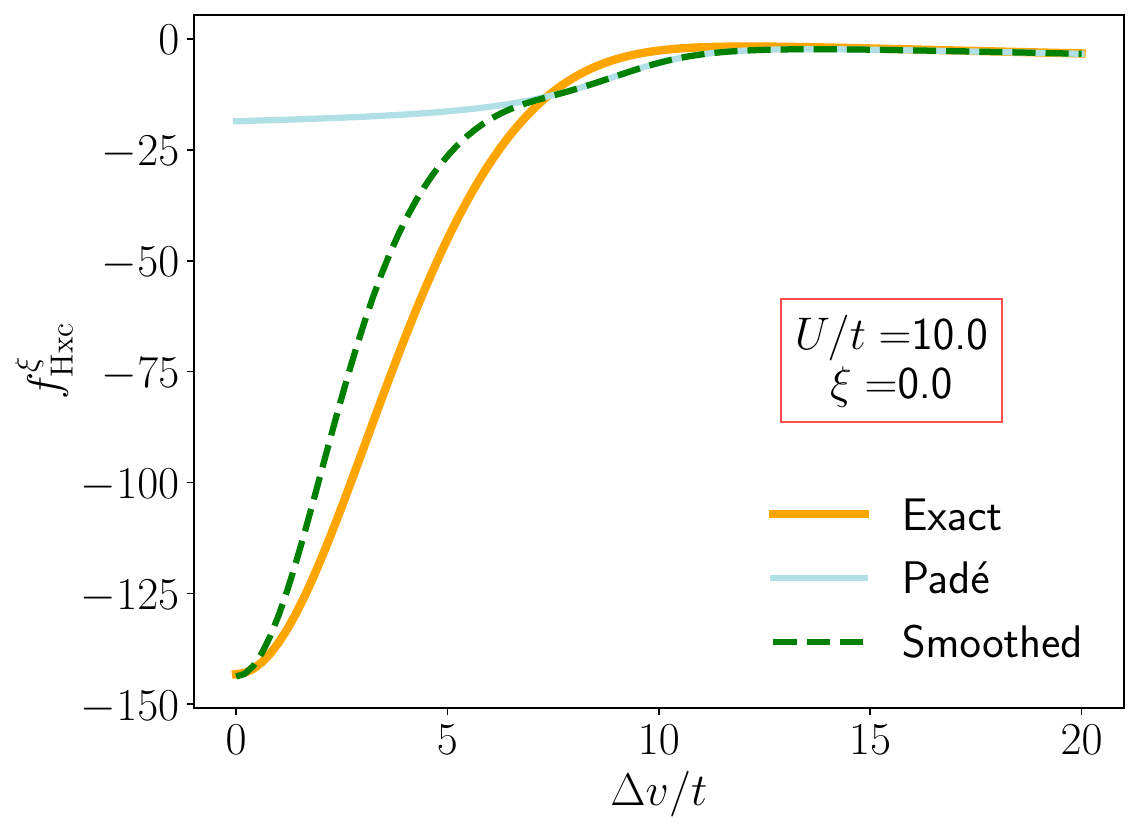}&
    \includegraphics[width=0.32\linewidth]{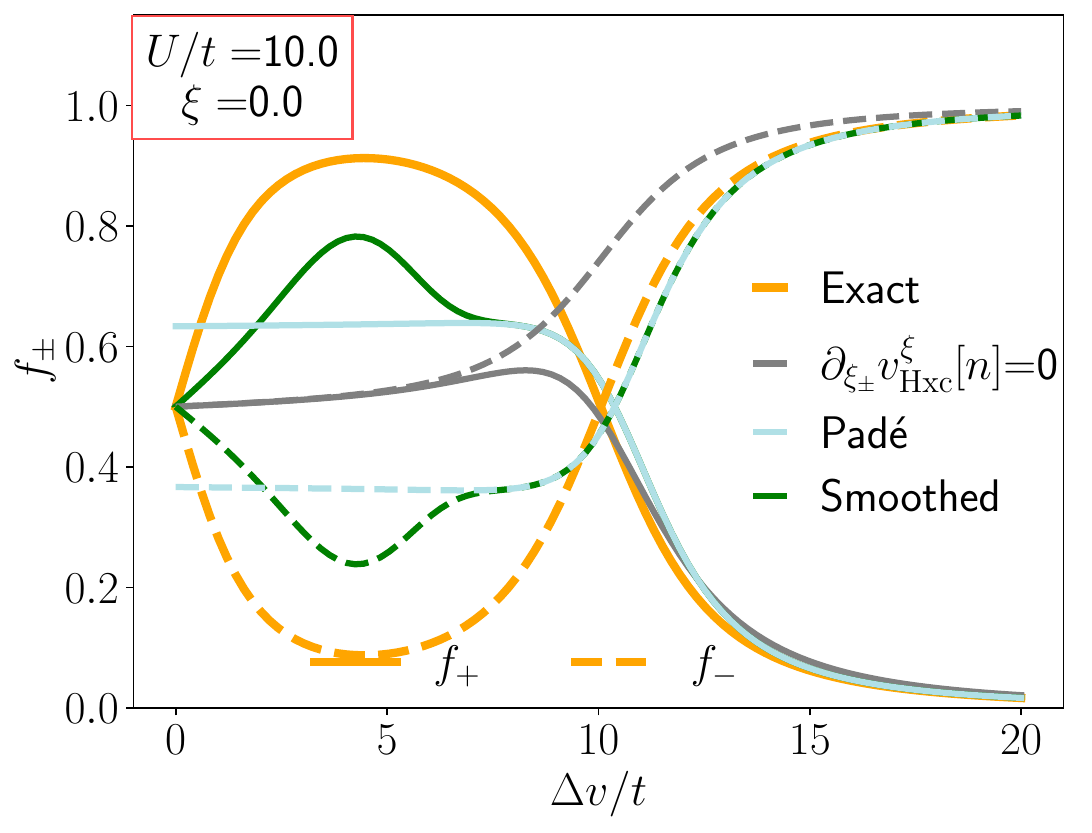}
    \end{tabular}
    \caption{Same as Fig.~\ref{fig:spvspfU5x0} with $U/t=10$.}
    \label{fig:spvspfU10x0}
\end{figure}

\begin{figure}[h]
    \centering
    \begin{tabular}{ccc}
    \includegraphics[width=0.32\linewidth]{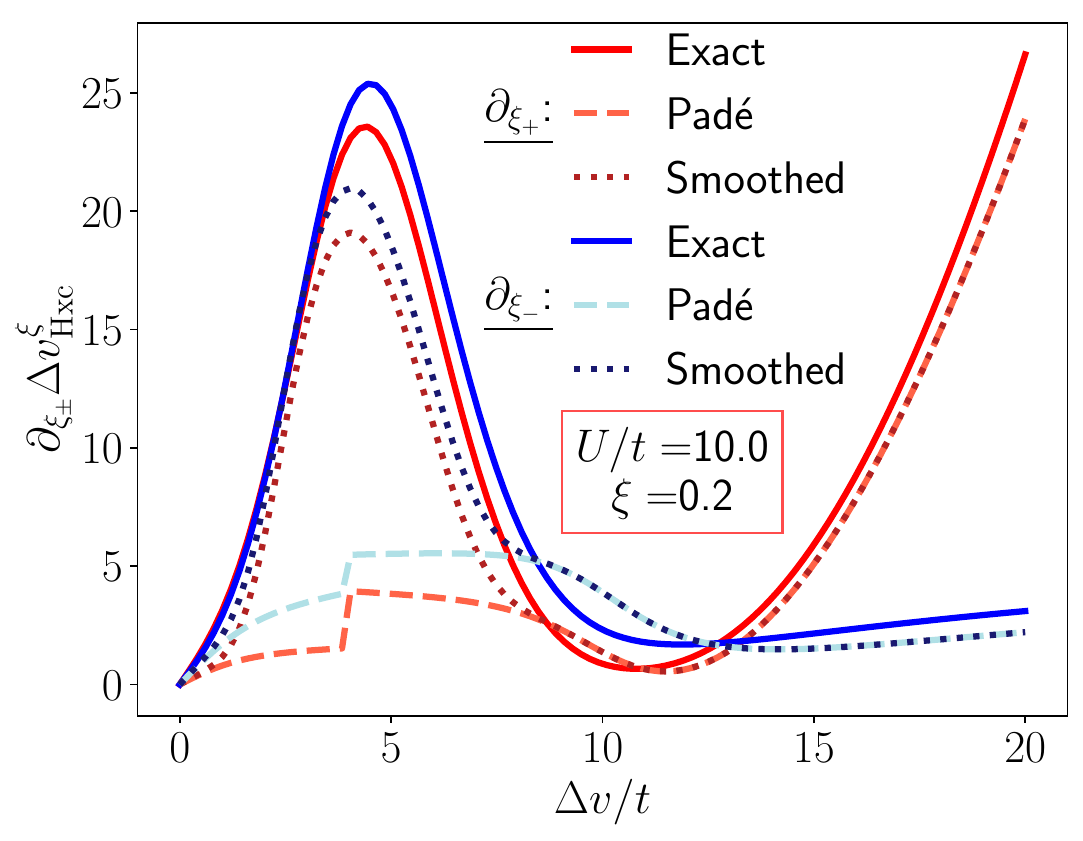}&
    \includegraphics[width=0.32\linewidth]{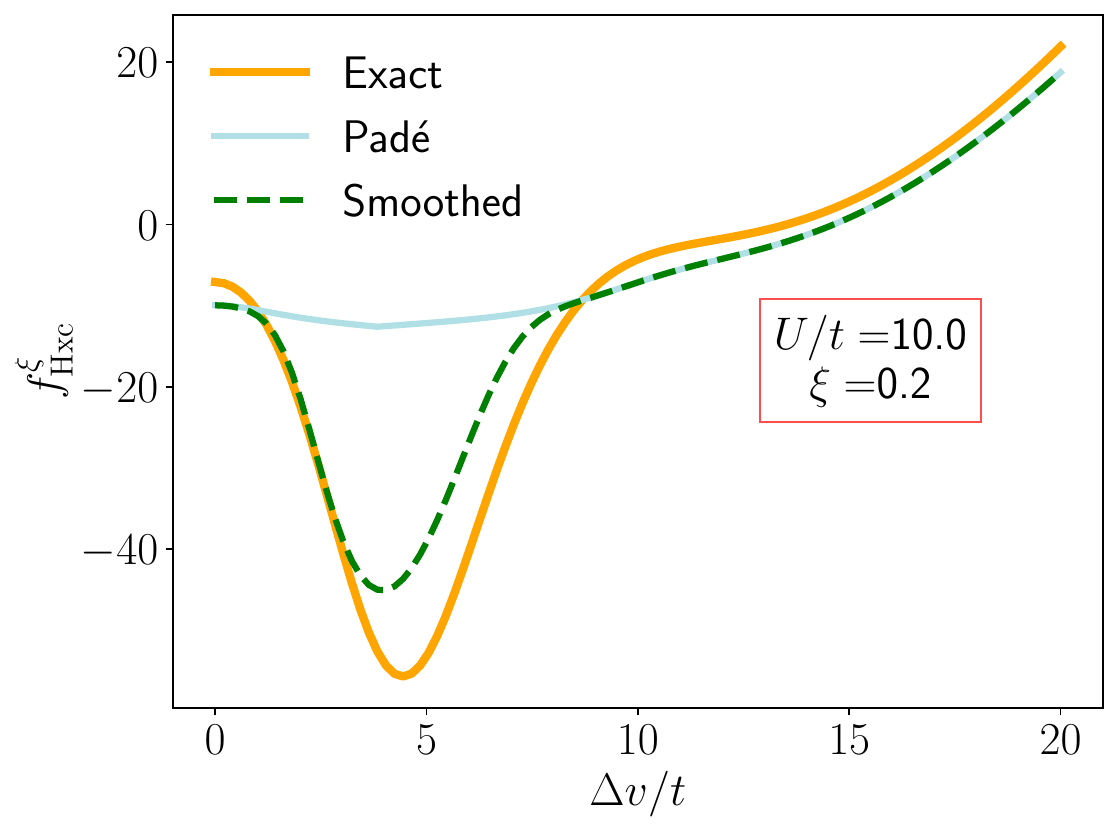}&
    \includegraphics[width=0.32\linewidth]{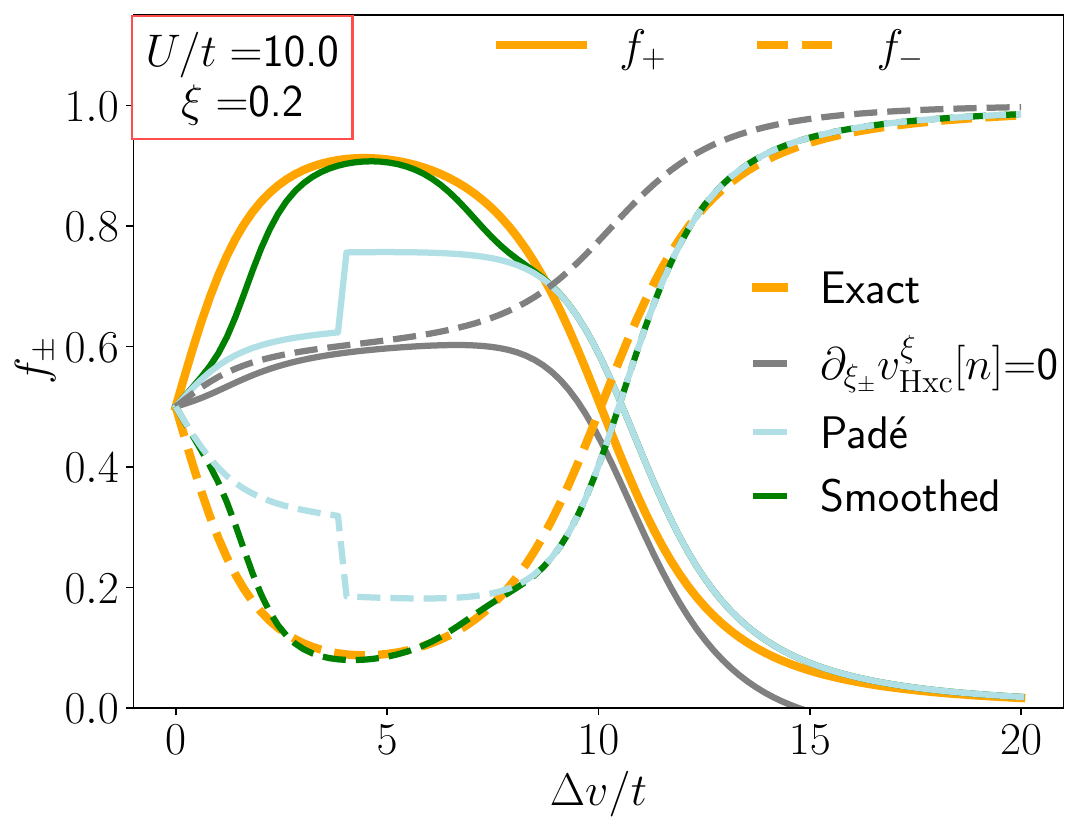}
    \end{tabular}
    \caption{Same as Fig.~\ref{fig:spvspfU5x0} with $U/t=10$ and $\xi=\xi_+=\xi_-=0.2$.}
    \label{fig:spvspfU10x2}
\end{figure}

Although $U \, \gamma^{\bxi}(n)$ defined through Eq.~\eqref{eqapp:pade2} is the exact limit of $E^{\bxi}_{\rm Hxc}(n)$ for $u\rightarrow\infty$, its discontinuous behavior is a poor approximation of the strong correlation regime (say, $U/t>5$.) On Figs.~\ref{fig:spvspfU5x0} to \ref{fig:spvspfU10x2}, the right panels show it does not reproduce Fukui functions accurately for $\Delta v < U$, considering to sets of weights and two values of $U$. The discontinuities in $\gamma^{\bxi}$ derivatives are reflected in the Fukui functions, at $\Delta v=0$ for the 0-weight case, and displaced around $2 \xi \, U$ for $\xi=\xi_+=\xi_-=0.2$. The left panels show the Padé approximant error comes from a missing contribution to the weight-derivative of the Hxc potential for low $\Delta v$. Its prediction for the Hxc kernel, shown in middle panels, is also poor in this range.
It is reasonable to assume that for high-but-finite $U$, the Hxc energy has continuous derivatives with respect to the density and weights, which yield cross-derivative contributions to $\frac{\partial v^{\bxi}_{\rm Hxc}(n)}{\partial\xi_{\pm}}|_{n=n^{\bxi}}$. Similarly, it would induce important variations of the kernel around the derivative discontinuities of the Padé approximant that we observe.

We thus replace the original expression for $\gamma^{\bxi}(n)$ by a smoothed version:
\begin{equation}
    \begin{split}
        \bar{\gamma}^{\bxi}(n) & = \eta(n-1) + \frac{1}{2}\left(\xi_+ - \xi_- + \frac{1}{k_{\xi}}\ell n(1+e^{k_{\xi}[\xi_+ + \xi_- - 2*\eta(n-1)]} \right)
            \\
            \eta(n-1) & = \frac{2}{k_n} \ell n \left(1+e^{k_n*(n-1)}\right) - (n-1)
    \end{split}
\end{equation}
where the parameters $k_n$ and $k_{\xi}$ control the stiffness of the function, and should be functions of $U$. This is an opportunity to explore how conditions on the functional in familiar regimes can guide DFA development. To do so, we set $k_n$ and $k_{\xi}$ so that $f^{\bxi}_{\rm Hxc}(n)$ is correctly reproduced in the symmetric case ($\Delta v=0$), corresponding to the uniform electron gas limit, for zero weights. This is illustrated on middle panels of Figs.~\ref{fig:spvspfU5x0} and~\ref{fig:spvspfU10x0}. Multiple couple values $(k_n,k_{\xi})$ can satisfy this requirement. As an additional constraint, we minimize the difference between the exact and Padé slopes of $\displaystyle \frac{\partial \Delta v^{\bxi}_{\rm Hxc}(n)}{\partial \xi_{\pm}}$ for $\Delta v=0$ and zero weights, as seen on left panels of aforementioned figures. This yields the parameters $(k_n=64,k_{\xi}=15)$ for $U=5$ and $(k_n=130,k_{\xi}=25)$ for $U=10$. The improvement over the Padé approximation with discontinuous derivatives can be seen on right panels of Figs.~\ref{fig:spvspfU5x0} and~\ref{fig:spvspfU10x0} for the prediction of Fukui functions. More interesting, in our opinion, is the impact outside the 0-weight limit. One can see the smoothed Padé approximant improves substantially the Hxc kernel, weight-derivatives of Hxc potential, and resulting Fukui functions for $\xi=\xi_+=\xi_-=0.2$ on Figs.\ref{fig:spvspfU5x2} and \ref{fig:spvspfU10x2}. This is a promising result for transferability as only the 0-weight limit was used to improve the Padé approximant. Of course, other constraints could be considered, and our choice of smoothing functions is entirely arbitrary. In that regard, performing a Taylor expansion in $1/u$ around $1/u=0$, in the spirit of ISI\cite{Gould2023_Electronic} would be informative in realistic systems. The present constructions simply serves to demonstrate care must be taken with discontinuous derivatives in ensemble DFAs.


\end{document}